\documentclass[fleqn,usenatbib,useAMS]{mnras}

\usepackage{graphicx}	
\usepackage{amsmath}	
\usepackage{amssymb}	
\usepackage{multicol}        
\usepackage{bm}		
\usepackage{pdflscape}	
\usepackage{epstopdf}
\usepackage{xcolor}

\newcommand{\mpc}{{\rm Mpc}}

\newcommand{\kms}{{\rm ~km~s^{-1}}}

\usepackage[T1]{fontenc}
\usepackage{ae,aecompl}

\title[CGs analysis using weak gravitational lensing
II]{Compact Groups analysis using weak gravitational lensing
 II: CFHT Stripe 82 data}

\author[Chalela et al.]{Mart\'in Chalela$^{1,2}$, Elizabeth Johana Gonzalez$^{1,2}$, Mart\'in Makler$^{3}$, \newauthor
Diego Garc\'ia Lambas$^{1,2}$,
Maria E. S. Pereira$^{4}$, Ana Laura O'Mill$^{1,2}$, Huan Yuan Shan$^{5}$ \\
$^{1}$ Instituto de Astronom\'{\i}a Te\'orica y Experimental (IATE-CONICET),
 Laprida 854, X5000BGR, C\'ordoba, Argentina.\\
$^{2}$ Observatorio Astron\'omico de C\'ordoba, Universidad Nacional de C\'ordoba, Laprida 854, X5000BGR, C\'ordoba, Argentina.\\
$^{3}$ 
Centro Brasileiro de Pesquisas F\'{\i}sicas, Rio de Janeiro, RJ 22290-180, Brasil\\
$^{4}$ Brandeis University, 415 South Street, Waltham, MA 02453, USA.\\
$^{5}$Argelander-Institut f\"ur Astronomie, Auf dem H\"ugel 71, 53121 Bonn, Germany. }

\pubyear{2018}

\begin{document}
\label{firstpage}
\pagerange{\pageref{firstpage}--\pageref{lastpage}}
\maketitle

\begin{abstract}
In this work we present a lensing study of Compact Groups (CGs) using data obtained from the high quality Canada-France-Hawaii Telescope Stripe 82 Survey. Using stacking techniques we obtain the average density contrast profile. We analyse the lensing signal dependence on the groups surface brightness and morphological content, for CGs in the redshift range $z = 0.2 - 0.4$. We obtain a larger lensing signal for CGs with higher surface brightness, probably due to their lower contamination by interlopers. Also, we find a strong dependence of the lensing signal on the group concentration parameter, with the most concentrated quintile showing a significant lensing signal, consistent with an isothermal sphere with $\sigma_V =336 \pm 28$ km/s and a NFW profile with $R_{200}=0.60\pm0.05$ $h_{70}^{-1}$Mpc. We also compare lensing results with dynamical estimates finding a good agreement with lensing determinations for CGs with higher surface brightness and higher concentration indexes. On the other hand, CGs that are more contaminated by interlopers show larger dynamical dispersions, since interlopers bias dynamical estimates to larger values, although the lensing signal is weakened.
\end{abstract}

\begin{keywords}
gravitational lensing: weak, galaxies: groups: general
\end{keywords}



\section{Introduction}
According to the standard cosmological model of structure formation ($\Lambda$CDM), less massive objects merge to form larger systems \citep[for a detailed revision see ][]{Kravtsov2012}. Clusters and superclusters of galaxies are therefore the largest, and most recently formed, gravitationally bound structures \citep{Voit2005}. In this scenario, galaxies group together in configurations ranging from galaxy pairs to rich clusters that could contain thousands of members. Given this hierarchical scenario of structure formation, the study of small galaxy systems such as compact groups, as well as the properties of their member galaxies, is essential to fully understand galaxy formation and evolution \citep{Bitsakis2010,Mendes1994,Verdes-Montenegro2001}.\\
It is well known that several properties of galaxies depend on the environment. \citet{Oemler1974} and \citet{Dressler1980} showed that galaxy morphology correlates with local density: in low density environments, galaxies tend to be blue, star-forming and late-type, while dense environments are dominated by red, early-type galaxies. Several processes are known to modify galaxy properties. For example, the hot intracluster gas is the main responsible for stripping the gas of galaxies in the core of clusters. Tidal interactions and mergers have a strong impact on morphology, transforming spiral galaxies into elliptical and S0s \citep{Toomre1972}. The extremely dense environment of CGs, where galaxies lie within a region of just a few galaxy radii with low velocity dispersions \citep[$\sim 200 \kms; e.g. $][]{McConnachie2009}, provides an ideal scenario to study galaxy merging and the impact of environment on galaxy evolution \citep[eg. ][]{Rubin1990,Rodrigue1995,Amram2007,Coziol2007,Plauchu-Frayn20101,Gallagher2010,Konstantopoulos2012,Sohn2013,Vogt2015}. In particular, \citet{Lee2017} find that galaxy evolution is faster in compact groups than in the central regions of clusters. Also, they obtain that mid-infrared colours of compact group early-type galaxies are on average bluer than those of cluster early-type galaxies \citep{Lee2017}. \\
\citet{Plauchu-Frayn20101} and \citet{Plauchu-Frayn20102} compare galaxies in CGs with isolated pairs of galaxies using optical and near-infrared observations, obtaining that stellar populations in the CGs galaxies are older and more dynamically relaxed. This is in agreement with the morphology-density relation, which suggests that CGs may have a high fraction of red, early-type galaxies. \citet{Coenda2012} studied several properties of galaxies in CGs, finding that they are, on average, smaller in size and have higher surface brightness than galaxies in the field or in loose groups. When analyzing the concentration index, i.e. $c=r_{90}/r_{50}$ (where $r_{90}$ and $r_{50}$ are the radii enclosing 90\% and 50\% of the $r$-band Petrosian flux, respectively), they found that galaxies are also more concentrated in CGs than in the field or in loose groups. This result is consistent with \citet{Strateva2001} where they suggest that the concentration index provides a suitable proxy to morphology, able to distinguish late and early-type galaxies with high confidence.\\
In \citet[][hereafter Paper\,I]{Chalela2017} we presented the first statistical weak lensing analysis of a sample of CGs from $z=0.06$ up to $z=0.2$, using stacking techniques in the Sloan Digital Sky Survey \cite[SDSS, ][]{York2000} images. Our results suggest that CGs mass distribution is well modelled by a Singular Isothermal Sphere profile with centre given by a galaxy luminosity-weighted scheme. We also found that groups with more concentrated galaxy members show steeper mass profiles and larger velocity dispersions. Given that galaxies in CGs are expected to be early-type on average, this result supports the idea that the concentration index could be used to asses the presence of interlopers. Therefore, our weak lensing analysis allowed us to infer the presence of gravitationally unbound systems given their weakening effect on the lensing signal. However, the uncertainties in our estimates did not allow us to confirm our results with high statistical confidence.\\
In this work, we apply a weak lensing analysis to a sample of CGs taken from \citet{McConnachie2009} in the redshift range $0.2<z<0.4$, complementing our previous study. Background galaxies are selected from the CFHT Stripe 82 Survey (CS82, Erben et al. in prep.). We take advantage of this deeper dataset provided by the Canada-France-Hawaii Telescope, designed with the goal of complementing existing Stripe 82 SDSS $ugriz$ imaging with high quality $i$-band imaging suitable for weak lensing measurements. This high quality data allows us to perform a detailed analysis of CGs mass dependence on astrophysical parameters. We also complement our analysis with photometric redshifts of background galaxies to achieve higher signal-to-noise in lensing measurements. \\
The paper is organized as follows: in Section \ref{sec:data} we describe the sample of CGs used for the analysis and the CS82 galaxy data. We provide details of the lensing analysis in Section \ref{sec:analysis}. In Section \ref{sec:results} we show the resulting lensing masses and analyze their dependence with CGs surface brightness and a weighted luminosity concentration parameter. We also compare our results with dynamical mass estimates. Finally, we discuss the main results in Section \ref{sec:conclusions}. Throughout this paper we adopt a standard cosmological model with parameters $H_{0} = 70\kms\mpc^{-1}$, $\Omega_{m} = 0.3$ and $\Omega_{\Lambda} = 0.7$.
\section{Data Selection}
\label{sec:data}
\subsection{CGs catalogue}
\begin{figure} 
\includegraphics[scale=0.45]{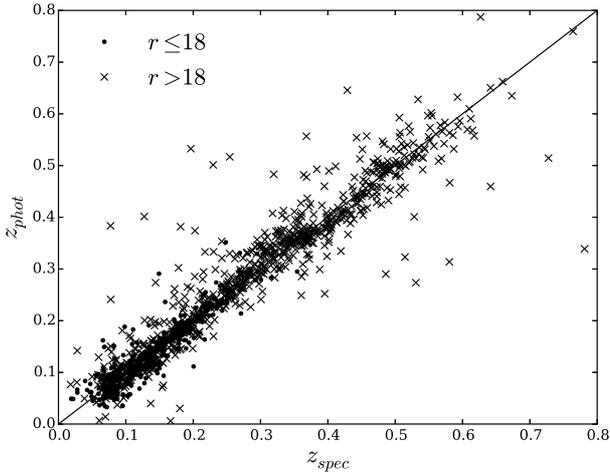}
\caption{Spectroscopic ($z_{spec}$) vs photometric ($z_{phot}$) redshifts of member galaxies with $r$ magnitudes larger than 18 (crosses) and lower than 18 (points). The solid line corresponds to the identity}
\label{fig:zphot} 
\end{figure}

\begin{figure}
\includegraphics[scale=0.45]{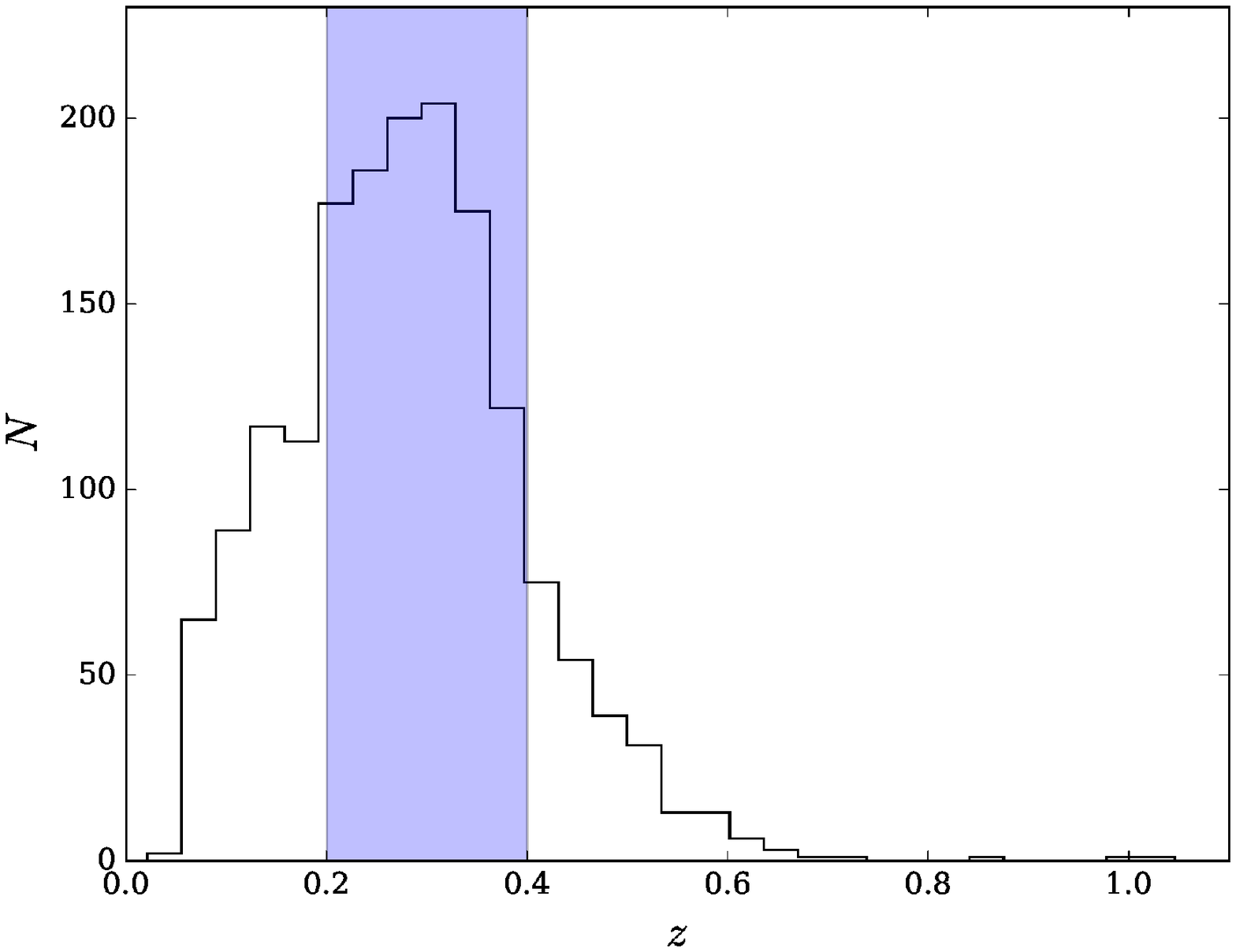}
\includegraphics[scale=0.45]{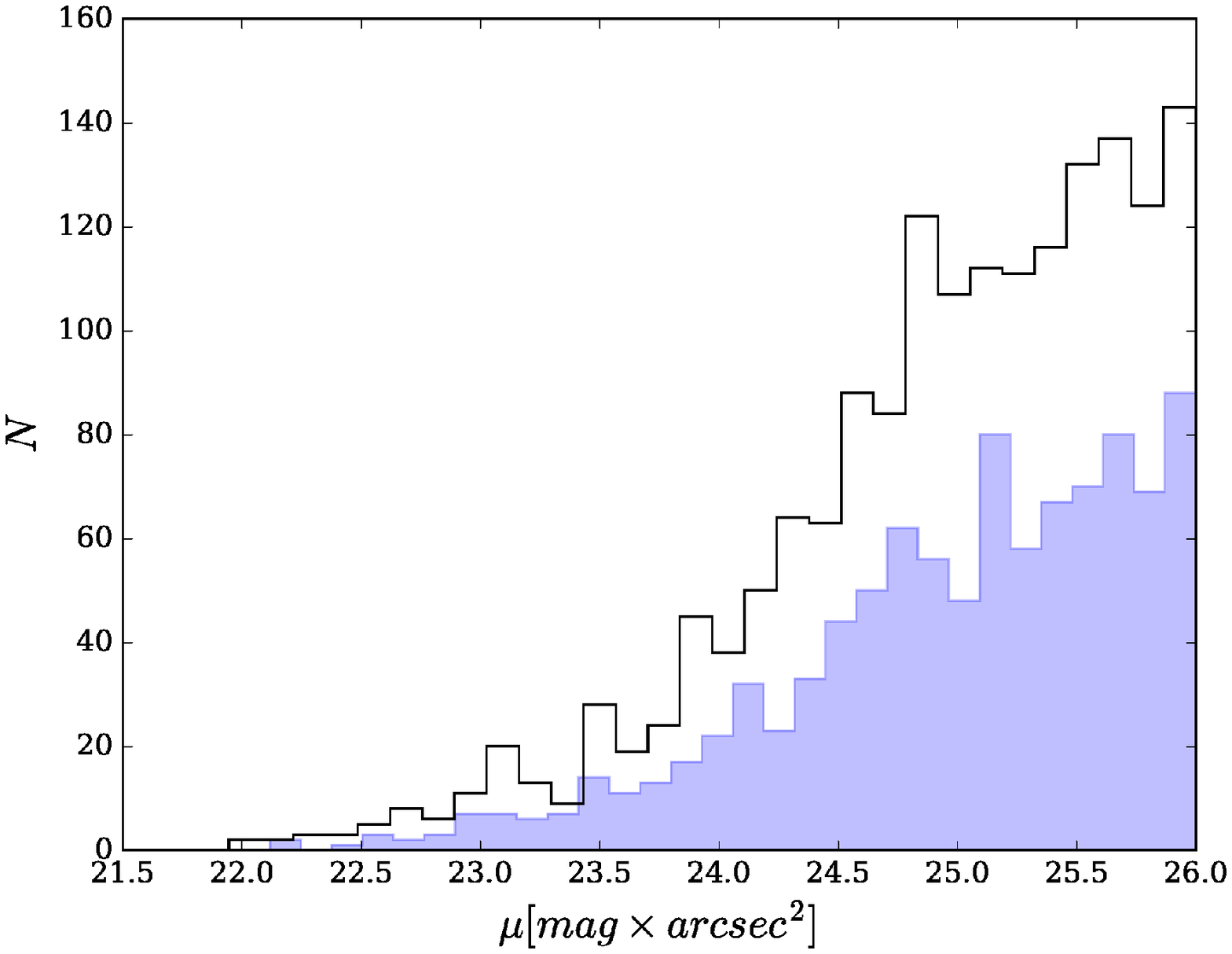}
\caption{Distributions of parameters of the CGs within the CS82 region: redshift (\textit{upper}) and surface brightness, $\mu$, (\textit{lower}) in magnitudes/arcsec$^2$. The shaded regions corresponds to the range of redshifts considered for the analysis and its resulting surface brightness distribution.}
\label{fig:groups}
\end{figure}

We consider a CG sample taken from Catalogue B of \citet{McConnachie2009} based on SDSS photometric data. These groups were selected according to the \citet{Hickson1982} criteria:
\begin{enumerate}
\item N($\Delta m = 3 $) $\geq 4$;
\item $\theta_N \geq 3 \theta_G$;
\item $\mu \leq 26.0 $ mag\,arcsec$^{-2}$
\end{enumerate}
where $N(\Delta m = 3)$ is the number of member galaxies within 3 magnitudes of the brightest galaxy, $\theta_G$ is the angular diameter of the smallest circle that enclose the centres of these galaxies, $\theta_N$ is the angular diameter of the largest concentric circle with no additional galaxy in this magnitude range or brighter, and $\mu$ is the effective surface brightness of member galaxies (where the total flux is averaged over the circle of angular diameter $\theta_G$). Catalogue B contains 74791 CGs with member galaxies in the magnitude range $14.5 \leq r \leq 21.0$. \\
Only 1694 CGs matched the CS82 region considering the masked areas. Redshifts of member galaxies (7153 galaxies) in this sample were updated with other spectroscopic redshift information from SDSS Data Release 12 \citep{Alam2015}, a compilation of spectroscopic data from \citet{Soo2017} and the 2dF catalogue \citep{Colless2003}. In order to increase our sample with redshift information we also used photometric redshifts computed by \citet{Reis2012}. To test the reliability of these redshift estimates, we use the sample of 1186 CG member galaxies with spectroscopic data. In Figure \ref{fig:zphot} we show $z_{phot}$ vs. $z_{spec}$ where it can be seen a very good agreement between spectroscopic and photometric redshift estimates, with $<(z_{spec}$-$z_{phot})^2>^{(1/2)}=0.06$. When we restrict the sample to galaxies with $r>18$ we obtain a rms value of $\sigma_{phot}=0.08$.\\
We compute the redshift of each group by averaging the spectroscopic redshifts of member galaxies. In the cases where the group has no galaxies with spectroscopic information (758 CGs), we use the weighted average of the available photometric redshifts according to the $z_{phot}$ uncertainties. Magnitudes of member galaxies in the selected systems were updated with DR12 photometric catalogues. Some objects previously classified as galaxies by the SDSS-DR6 automatic pipelines were reclassified as a different object in DR12 (for example stars or wrongly de-blended galaxies). We restrict to those CGs with all their members still classified as galaxies. In order to study CGs with similar intrinsic properties, we restrict our sample to systems within $\sim 1 \sigma$ around the mean of the distribution of CG in redshift, which yields the interval $0.2 < z < 0.4$. Restricting to this redshift range, the total sample consists of 975 CGs. In Figure \ref{fig:groups} we show the distribution of redshifts and surface brightness of all CGs within the the CS82 region together with the selected sample.   
\subsection{Weak lensing data}
The CS82 survey is a joint Canada-France-Brazil project that was designed with the goal of complementing existing Stripe\,82 SDSS $ugriz$ imaging with high quality $i'$-band imaging suitable for weak and strong lensing measurements (Erben et al. in prep.). CS82 is built from 173 MegaCam $i'$-band images and corresponds to an area of 160 degrees$^2$ (129.2 degrees$^2$ after masking out bright stars and other image artifacts). The Point Spread Function (PSF) for CS82 varies between $0.4''$ and $0.8''$ over the entire survey with a median of $0.6''$. The limiting magnitude of the survey is $ i' \sim 24$ \citep{Leauthaud2017}. Shear catalogues were constructed using the same weak lensing pipeline developed by the CFHTLenS collaboration using the \textit{lens}fit Bayesian shape measurement method \citep{Miller2013}. Shear calibration factors and tests of systematics were applied in the same way as in \citet{Heymans2012}. For each galaxy, an additive calibration parameter $c_2$ is applied to the $e_2$ measured ellipticity component. Also a multiplicative shear calibration factor, $m(\nu_{SN},l)$, is considered, where $\nu_{SN}$ is the signal-to-noise ratio and $l$ is the size of the source.\\
In order to perform the lensing analysis, we select the galaxies by applying the following cuts to the \textit{lens}fit output catalogue: MASK $\leq$ 1, FITCLASS = 0, $i' < 24.7$, and $w > 0$.  Here FITCLASS is a flag parameter given by \textit{lens}fit to select sources classified as galaxies, $w$ is a weight parameter which takes into account errors on the shape measurement and the intrinsic shape noise, that ensures that galaxies have well-measured shapes \citep[see details in ][]{Miller2013} and MASK is a masking flag. We select galaxies using the centres given by \citet{McConnachie2008} within a $0.5$deg radius. Background galaxies, i.e. galaxies affected by the lensing effect, are identified using the \citet{Soo2017} photometric redshift catalogue. We select galaxies with Z\_BEST > $z_g$+0.1 and ODDS\_BEST $\geq 0.6$, where $z_g$ is the CG redshift, Z\_BEST is the photometric redshift provided by \citet{Soo2017} and ODDS\_BEST expresses the quality of the redshift estimates, from 0 to 1.
\section{Weak lensing analysis}

\label{sec:analysis}
Since CGs are low mass systems ($\sim 10^{13} M_\odot$) it is necessary to perform a stacking analysis in order to obtain a reliable lensing signal. We briefly describe the stacking technique as detailed in \citet{Eli2016}.\\
Gravitational lensing effects are characterized by the convergence $\kappa$, which reflects the isotropic stretching of source images, and an anisotropic distortion given by the complex-value lensing shear, $\gamma = \gamma_1 + i \gamma_2$. Using the second derivative of the projected gravitational potential to express the shear and convergence, it can be shown that for a lens with an axially-symmetric projected mass distribution, the tangential component of the shear, $\gamma_T$, is related to the surface mass density, $\Sigma(r)$, through \citep{Bartelmann1995}:
\begin{equation}
\label{Dsigma}
\tilde{\gamma}_{T}(r) \times \Sigma_{crit} = \bar{\Sigma}(<r) - \bar{\Sigma}(r) \equiv \Delta\tilde{\Sigma}(r),
\end{equation}
where we define the density contrast, $\Delta\tilde{\Sigma}$. Here $\tilde{\gamma}_{T}(r)$ is the averaged tangential component of the shear in a ring of radius $r$, $\bar{\Sigma}(<r) $ and $\bar{\Sigma}(r)$ are the average projected mass distribution within a disk and in a ring of radius $r$, respectively. On the other hand, the cross-component of the shear, $\gamma_{\times}$, defined as the component tilted at $\pi$/4 relative to the tangential component, should be zero. In the weak-lensing regime, the image of a circular source appears elliptical and the induced ellipticity can be directly related to the shear, $\gamma \approx e$. Here we define the ellipticity as a complex number, $e = e_1 + i e_2$, with magnitude $\mid e \mid = (a-b)/(a+b)$ and orientation angle determined by the direction of the major axis, $a$. If the source has an intrinsic ellipticity $e_s$ the observed ellipticity will be $e \approx e_s + \gamma$ \citep{Bartelmann2001}. If we consider many sources with intrinsic ellipticities randomly orientated so that $\langle e_s \rangle = 0$, the ensemble average ellipticity after lensing gives an unbiased estimate of the shear: $\langle e \rangle \approx \gamma$. Therefore, we can estimate the shear by averaging the shape of background sources affected by the lens effect.\\
When considering a sample of lensing systems, the density contrast is obtained as the weighted average of the tangential ellipticity of background galaxies of the lens sample:
\begin{equation}
\langle \Delta \tilde{\Sigma}(r) \rangle = \frac{\sum_{j=1}^{N_{Lenses}} \sum_{i=1}^{N_{Sources,j}} \omega_{LS,ij} \times e_{T,ij} \times \Sigma_{crit,ij}}{\sum_{j=1}^{N_{Lenses}} \sum_{i=1}^{N_{Sources,j}} \omega_{LS,ij}},
\end{equation}
where $\omega_{LS,ij}$ is the inverse variance weight computed according to the weight, $\omega_{ij}$, given by $lens$fit algorithm for each background galaxy, $\omega_{LS,ij}=\omega_{ij}/\Sigma^2_{crit}$. $N_{Lenses}$ is the number of lensing systems and $N_{Sources,j}$ the number of background galaxies located at a distance $r \pm \delta r$ from the $j$th lens. $\Sigma_{crit,ij}$ is the critical density for the $i$th source of the $j$th lens, defined as:
\begin{equation}
\Sigma_{crit,ij} = \dfrac{c^{2}}{4 \pi G} \dfrac{D_{OS,i}}{D_{OL,j} D_{LS,ij}}.
\end{equation}
Here $D_{OL,j}$, $D_{OS,i}$ and $D_{LS,ij}$ are  the angular diameter distances from the observer to the $j$th lens, from the observer to the $i$th source and from the $j$th lens to the $i$th source, respectively. These distances are computed according to the adopted redshift for each CG (lens) and for each background galaxy (source). $G$ is the gravitational constant and $c$ is the light speed.\\
Once the density contrast is computed, we take into account a noise bias factor correction as suggested by \citet{Miller2013}, which considers the multiplicative shear calibration factor $m(\nu_{SN},l)$ provided by \textit{lens}fit. For this correction we compute:
\begin{equation}
1+K(z_L)= \frac{\sum_{j=1}^{N_{Lenses}} \sum_{i=1}^{N_{Sources,j}} \omega_{LS,ij} (1+m(\nu_{SN,ij},l_{ij}))}{\sum_{j=1}^{N_{Lens}} \sum_{i=1}^{N_{Sources,j}} \omega_{LS,ij}}
\end{equation}
following \citet{Velander2014,Hudson2015,Shan2017,Leauthaud2017,Pereira2018}. Then, we calibrate the lensing signal as:
\begin{equation}
\langle \Delta \tilde{\Sigma}^{cal}(r) \rangle = \frac{ \langle \Delta \tilde{\Sigma}(r) \rangle }{1+K(z_L)}.
\end{equation}
According to this equation, the projected density contrast profile is computed using non-overlapping concentric logarithmic annuli to preserve the signal-to-noise ratio of the outer region, from $r_{in}=170\,h^{-1}_{70}$\,kpc (where the signal becomes significantly positive) up to $r_{out} =2.5\,h^{-1}_{70}$\,Mpc. We adopt a luminosity weighted centre, taking into account the results presented in Paper\,I. We have considered different logarithm bins, choosing between a total of 7 and 9 radial bins.  We have not observed differences in the density profile parameters, within their uncertainties,  among these choices, therefore we fix the binning in order to obtain the lowest $\chi^2$ value. Given that the uncertainties in the estimated lensing signal are expected to be dominated by shape noise, we do not expect a noticeable covariance between adjacent radial bins and so we treat them as independent in our analyses. Accordingly, we compute error bars in the profile by bootstrapping the lensing signal using 100 realizations. We have tested several numbers for the bootstrap realizations (50, 100, 500,1000) and we have not found any significant variation of the errors in agreement with \citet{Vitorelli2018}.\\
Density contrast profiles are modelled considering only the main halo component using two mass models: a singular isothermal sphere (SIS) and a NFW profile \citep{Navarro1997}. We limit the profile analysis up to $2.5\,h^{-1}_{70}$\,Mpc in order to avoid the 2-halo term in the modelling. An initial estimation suggests that the relative amplitude of the 2-halo signal reaches $\sim20\%$ at this radius. The SIS profile is the simplest density model for describing a relaxed massive sphere with a constant isotropic one dimensional velocity dispersion, $\sigma_V$. At galaxy scales, dynamical studies \citep[eg.,][]{Sofue2001}, as well as strong \citep[eg.,][]{Davis2003} and weak \citep[eg.,][]{Brimioulle2013} lensing observations, are consistent with a mass profile following approximately an isothermal law. The shear ($\gamma_{\theta}$) and the convergence ($\kappa_{\theta}$) at an angular distance $\theta$ from the lensing system centre are directly related to $\sigma_V$ by:
\begin{equation}
\kappa_{\theta} = \gamma_{\theta} = \dfrac{\theta_{E}}{2 \theta}
\end{equation}
where $\theta_{E}$ is the critical Einstein radius defined as:
\begin{equation}
\theta_{E} = \dfrac{4 \pi \sigma_{V}^{2}}{c^{2}} \frac{D_{LS}}{D_{OS}}.
\end{equation}
Therefore,
\begin{equation}
\tilde{\Sigma}(\theta, \sigma_V) =  \dfrac{\sigma_{V}^{2}}{2 G \theta D_{OL}}. 
\end{equation}
We compute the mass within $R_{200}$ (the radius that encloses a mean density equal to 200 times the critical density of the Universe), \mbox{$M_{200}=200\rho_{crit}(z)\dfrac{4}{3}\pi\,R_{200}^{3}$}, following \citet{Leonard2010}:
\begin{equation}\label{eq:MSIS}
M_{200} =  \dfrac{2 \sigma_{V}^{3} }{\sqrt{50} G H(z)}, 
\end{equation} 
where $H(z)$ is the redshift dependent Hubble parameter and $G$ is the gravitational constant. \\
We also use the spherically symmetric  NFW profile \citep{Navarro1997} which depends on two parameters, the virial radius, $R_{200}$, and a dimensionless concentration parameter, $c_{200}$. This density profile is given by:
\begin{equation}
\rho(r) =  \dfrac{\rho_{crit} \delta_{c}}{(r/r_{s})(1+r/r_{s})^{2}}, 
\end{equation}
where $r_{s}$ is the scale radius, $r_{s} = R_{200}/c_{200}$ and $\delta_{c}$ is the cha\-rac\-te\-ris\-tic overdensity of the halo:
\begin{equation}
\delta_{c} = \frac{200}{3} \dfrac{c_{200}^{3}}{\ln(1+c_{200})-c_{200}/(1+c_{200})}.  
\end{equation}
Lensing formulae for the NFW density profile were taken from \citet{Wright2000}. If both, $R_{200}$ and $c_{200}$, are taken as free parameters to be fitted, we obtain significantly large uncertainties in $c_{200}$ values. This is due to the lack of information on the mass distribution near the lens centre and only a combination of strong and weak lensing can break this degeneracy. To overcome this problem, we follow \citet{Uitert2012,Kettula2015} and \citet{Pereira2018}, by using a fixed mass-concentration relation $c_{200}(M_{200},z)$, derived from simulations by \citet{Duffy2008}: 
\begin{equation}
c_{200}=5.71\left(M_{200}/2 \times 10^{12} h^{-1}\right)^{-0.084}(1+z)^{-0.47},
\end{equation}
where we take $z$ as the mean redshift value of the lens sample. The particular choice of this relation does not have a significant impact on the final mass values, which have uncertainties dominated by the noise of the shear profile.\\
To derive the parameters of each mass model profile we perform a standard $\chi^{2}$ minimization:
\begin{large}
\begin{equation}
\chi^{2} = \sum^{N}_{i} \dfrac{(\langle \tilde{\Sigma}^{cal}(r_{i})  \rangle - \tilde{\Sigma}(r_{i},p))^{2}}{\sigma^{2}_{\Delta \tilde{\Sigma}}(r_{i})},
\end{equation}
\end{large}\\
where the sum runs over the $N$ radial bins of the profile and the model prediction $p$ refers to either $\sigma_{V}$ for the SIS profile, or $R_{200}$ in the case of the NFW model. Errors in the best-fitting parameters are computed according to the variance of the parameter estimates.

\section{Results}
\label{sec:results}
\begin{figure}
\includegraphics[scale=0.4]{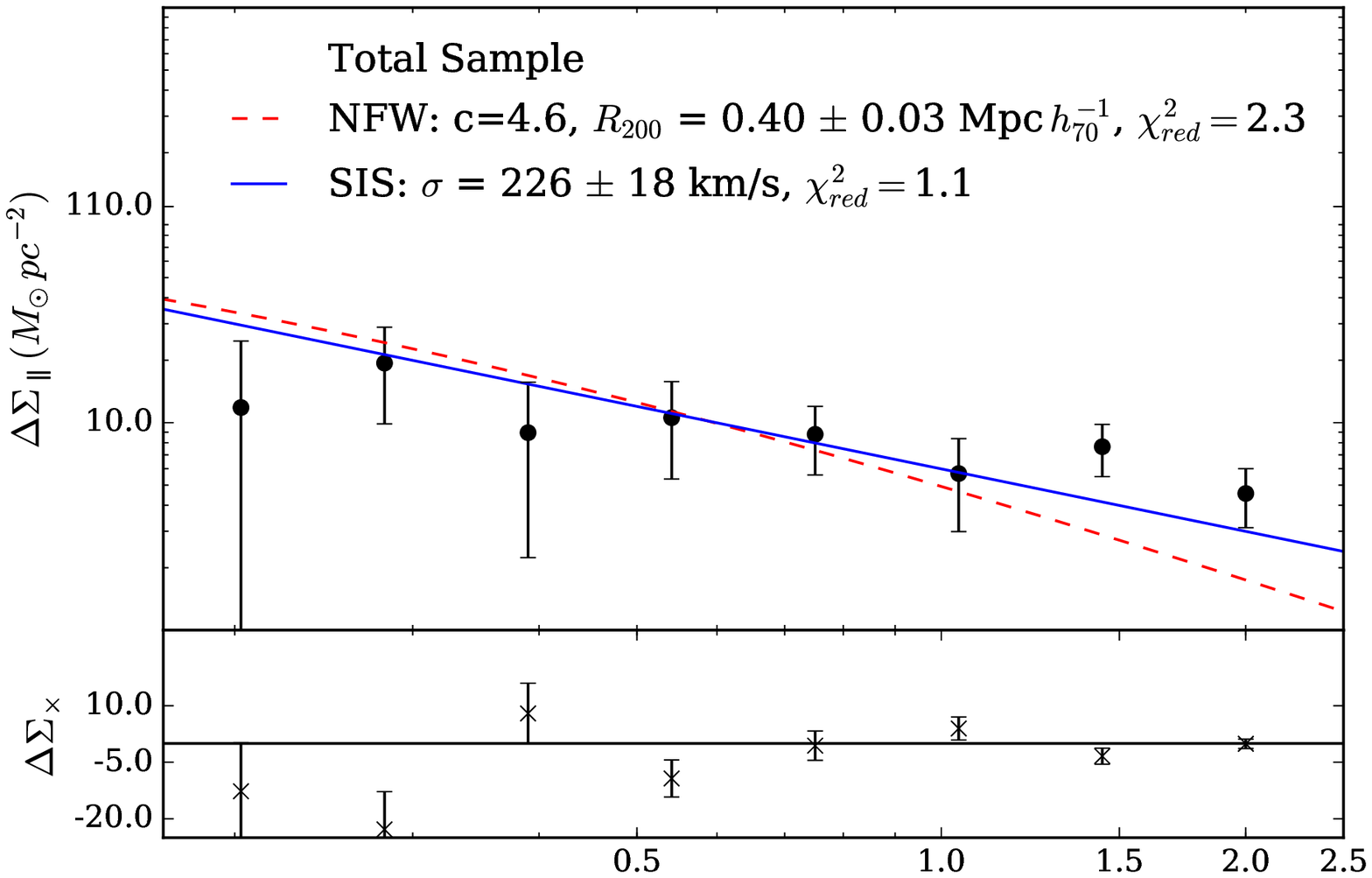}
\includegraphics[scale=0.4]{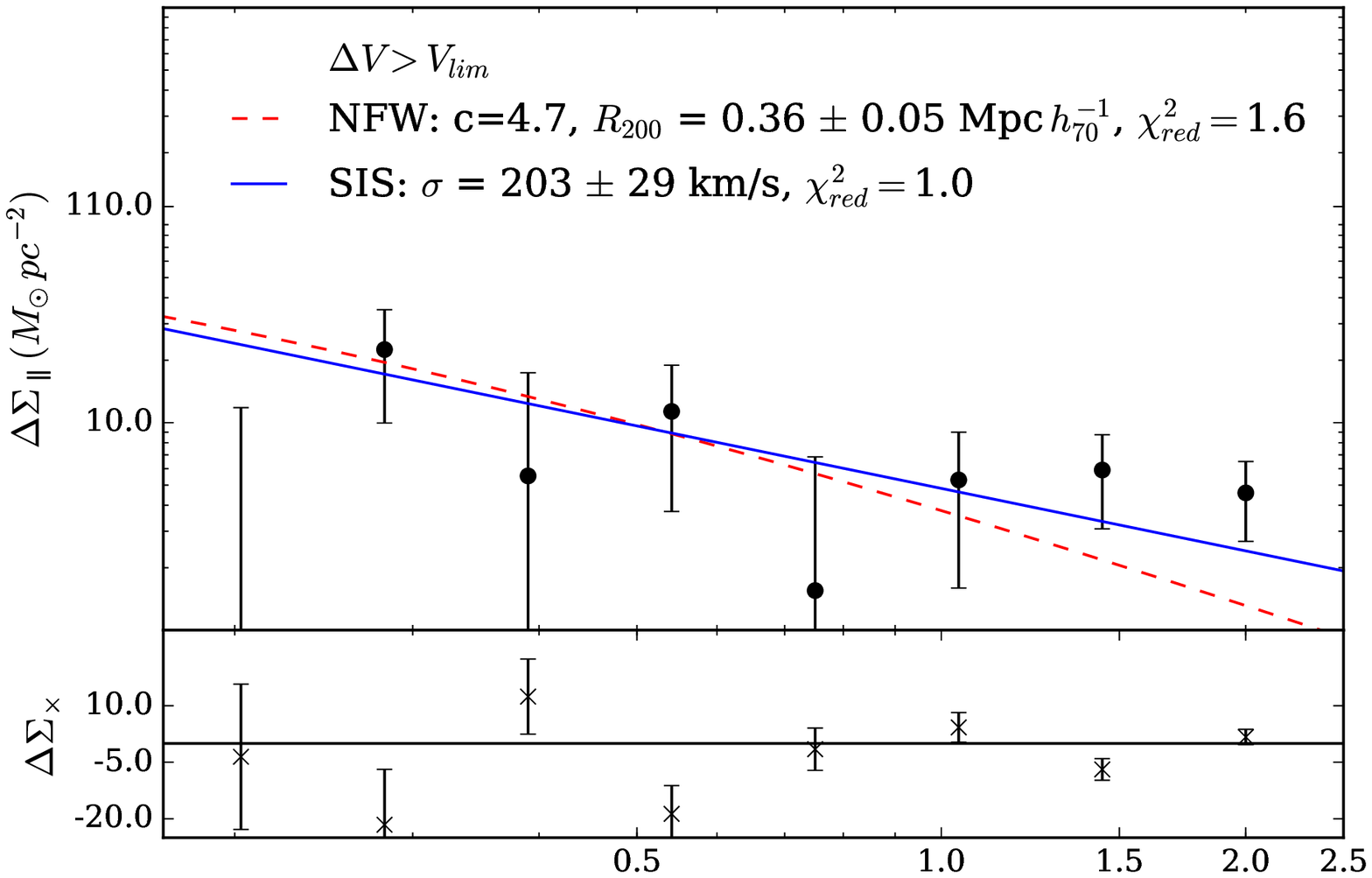}
\includegraphics[scale=0.4]{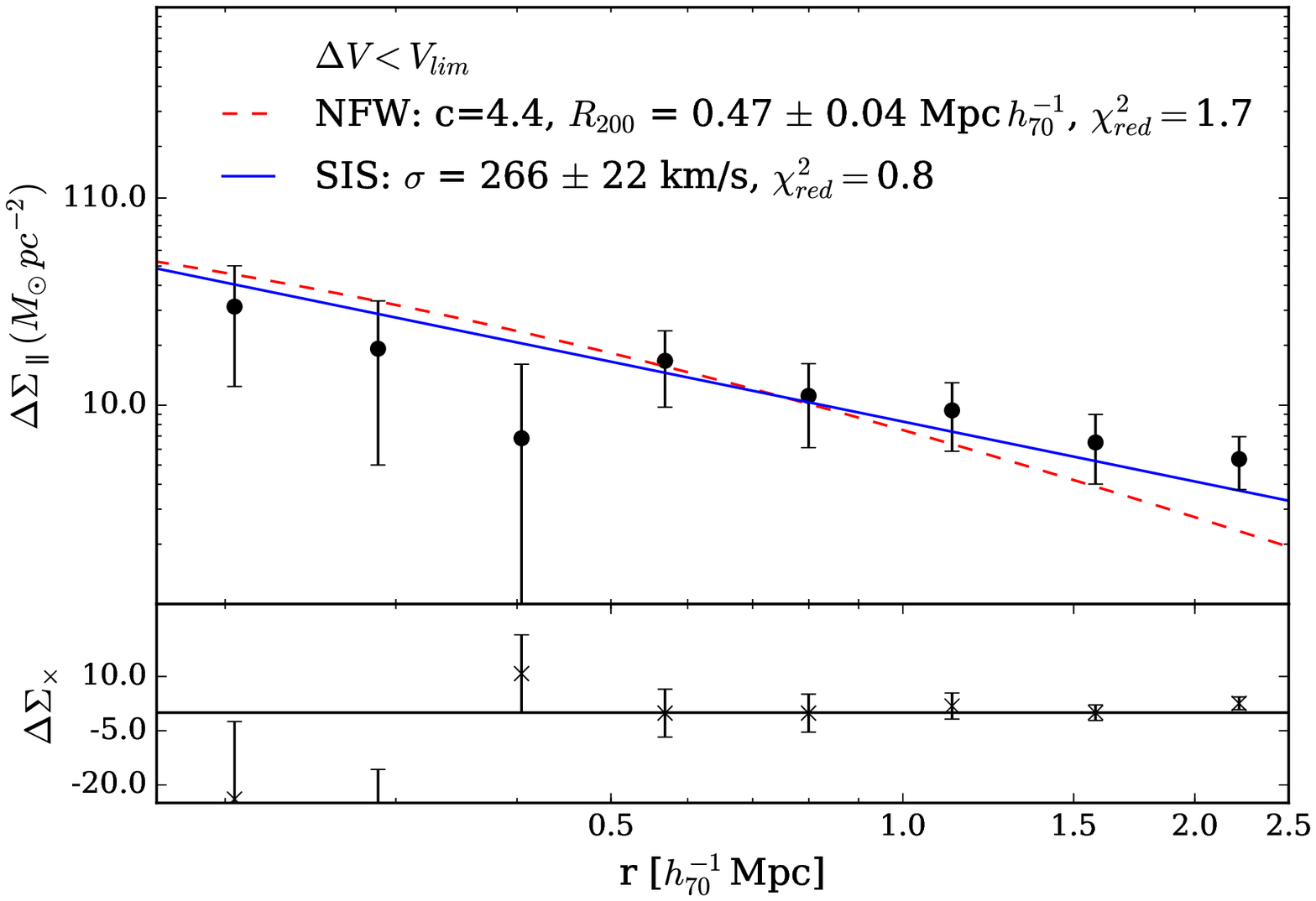}
\caption{Average density contrast profiles with their respective fitted models. From top to bottom: complete CGs sample, $\Delta V > V_{lim}$, $\Delta V < V_{lim}$. Results are summarized in Table \ref{table:results}.}
\label{fig:sample}
\end{figure}

\subsection{Analysis of different subsamples}
\label{subsec:zcutsample}
We first compute the density contrast profile for the total
sample (975 CGs), obtaining consistency with a SIS with $\sigma = 226$km\,s$^{-1}$ (see Figure \ref{fig:sample}). Since the presence of CGs in the sample that are not gravitationally bound systems can dilute the lensing signal, it is necessary to analyse different selection criteria, such as available redshifts, surface brightness, $\mu$, and concentration index of member galaxies. In this section we analyse the dependence of the lensing signal on these parameters. The results of the analysed subsamples are summarised in Table \ref{table:results}. We highlight the results for which the lensing masses are well determined (values have a confidence of at least 2.5$\sigma$, i.e., inconsistent with no signal).

\subsubsection{$\Delta V$ criterion}

In order to assess the impact of interlopers in CGs we divide the sample using a limiting relative velocity criterion. We consider those groups with $\Delta V < V_{lim}$ and those with $\Delta V > V_{lim}$, where: 
\begin{equation}
\Delta V = \max (c \mid z_i - z_j \mid /(1 + z_{ij} )), 
\end{equation}
with $z_{ij} = (z_i + z_j )/2$ and  $i \neq j$. The indexes $i$ and $j$ run over the total number of member galaxies with redshift estimation. $V_{lim} = 1000$km\,s$^{-1}$ when $z_i$ and $z_j$ are both spectroscopic estimates \citep[which is comonly used to clean the sample of interlopers, eg. ][]{McConnachie2009} and $V_{lim} = 2\sigma_{phot}c$ when $z_i$ and/or $z_j$ are photometric estimates. \\
In Figure \ref{fig:sample} we show the profiles for the total sample and for groups with $\Delta V < V_{lim}$ and with $\Delta V > V_{lim}$.  As it can be noticed the total sample and $\Delta V < V_{lim}$ show significant lensing signal. Also, applying the $\Delta V < V_{lim}$ criterion show considerable larger masses (twice those obtained from the $\Delta V > V_{lim}$ criterion), reinforcing that this sample is less affected by interlopers. It is also important to highlight that the $\Delta V < V_{lim}$ subsample could still include groups with interlopers since only a small fraction of the member galaxies have spectroscopic redshift estimates (17\% of member galaxies). On the other hand, the $\Delta V > V_{lim}$ subsample could comprise gravitationally bound groups including at least one interloper.

\begin{table*}
\caption{CGs results for the complete photometric sample}\label{tab:esp}
\begin{tabular}{@{}crccccccr@{}}
\hline
\hline
\rule{0pt}{1.05em}%
  Selection criteria   &   $N_{Lenses}$  & \multicolumn{3}{c}{SIS} & \multicolumn{3}{c}{NFW}  \\
 &        &$\sigma_{V}$ & $M_{200}$ & $\chi^2_{red}$ & $R_{200}$ & $M_{200}$ & $\chi^2_{red}$  \\
 &          & [km\,s$^{-1}$] &  [$10^{12} h_{70}^{-1} M_{\odot} $] & & [$h_{70}^{-1}$\,Mpc] & [$10^{12} h_{70}^{-1} M_{\odot} $] & \\

 \hline
\rule{0pt}{1.05em}%

\textbf{Total Sample} & \textbf{975} & $\mathbf{226 \pm 18}$ &$\mathbf{9.5 \pm 2.3}$ & $\mathbf{1.1}$ & $\mathbf{0.40 \pm 0.03}$ & $\mathbf{9.6 \pm 2.2}$ & $\mathbf{2.3}$ \\ 
$\Delta V > V_{lim}$ & 489 & $203 \pm 29$ &$6.8 \pm 3.0$ & $1.0$ & $0.36 \pm 0.05$ & $7.0 \pm 2.9$ & $1.6$ \\ 
$\mathbf{\Delta V < V_{lim}}$ & \textbf{486} & $\mathbf{266 \pm 22}$ &$\mathbf{15 \pm 4}$ & $\mathbf{0.8}$ & $\mathbf{0.47 \pm 0.04}$ & $\mathbf{16 \pm 4}$ & $\mathbf{1.7}$ \\ 
$\mathbf{Q1_{\mu}- \mu < 24.42} $ & \textbf{194} & $\mathbf{316 \pm 29}$ &$\mathbf{26 \pm 7}$ & $\mathbf{0.3}$ & $\mathbf{0.56 \pm 0.05}$ & $\mathbf{27 \pm 7}$ & $\mathbf{0.6}$ \\ 
$\mathbf{Q2_{\mu} - 24.42 \leq \mu < 24.89}$ & \textbf{195} & $\mathbf{269 \pm 35}$ &$\mathbf{16 \pm 6}$ & $\mathbf{0.4}$ & $\mathbf{0.48 \pm 0.06}$ & $\mathbf{17 \pm 6}$ & $\mathbf{0.6}$ \\ 
$Q3_{\mu} - 24.89 \leq \mu < 25.33$ & 195 & $231 \pm 39$ &$10 \pm 5$ & $0.7$ & $0.40 \pm 0.07$ & $10 \pm 5$ & $0.9$ \\ 
$Q4_{\mu} - 25.33 \leq \mu < 25.67$ & 195 & $148 \pm 65$ &$2.6 \pm 3.5$ & $0.5$ & $0.25 \pm 0.12$ & $2.4 \pm 3.5$ & $0.6$ \\ 
$Q5_{\mu}- \mu \geq 25.67$ & 195 & -            &     -        &   -   &       -         & -             & - \\          
$\mathbf{Q1_{\mu}}$ \textbf{and} $\mathbf{Q2_{\mu}}$ & \textbf{389} & $\mathbf{298 \pm 21}$ &$\mathbf{22 \pm 5}$ & $\mathbf{0.7}$ & $\mathbf{0.54 \pm 0.04}$ & $\mathbf{23 \pm 5}$ & $\mathbf{1.2}$ \\ 
$Q3_{\mu}$ to $Q5_{\mu}$ & 585 & $179 \pm 29$ &$4.7 \pm 2.3$ & $1.1$ & $0.31 \pm 0.05$ & $4.4 \pm 2.3$ & $1.7$ \\ 
$\mathbf{\Delta V < V_{lim}}$, $\mathbf{Q1_{\mu}}$ \textbf{and} $\mathbf{Q2_{\mu}}$ & \textbf{194} & $\mathbf{330 \pm 27}$ &$\mathbf{30 \pm 7}$ & $\mathbf{0.3}$ & $\mathbf{0.59 \pm 0.05}$ & $\mathbf{32 \pm 7}$ & $\mathbf{0.5}$ \\ 
$\Delta V < V_{lim}$, $Q3_{\mu}$ to $Q5_{\mu}$ & 291 & $190 \pm 42$ &$6 \pm 4$ & $0.6$ & $0.33 \pm 0.07$ & $6 \pm 4$ & $0.8$ \\ 
$\mathbf{\Delta V > V_{lim}}$, $\mathbf{Q1_{\mu}}$ \textbf{and} $\mathbf{Q2_{\mu}}$ & \textbf{195} & $\mathbf{282 \pm 31}$ &$\mathbf{18 \pm 6}$ & $\mathbf{0.7}$ & $\mathbf{0.50 \pm 0.05}$ & $\mathbf{19 \pm 6}$ & $\mathbf{0.9}$ \\ 
$\Delta V > V_{lim}$, $Q3_{\mu}$ to $Q5_{\mu}$ & 293 & - & - & - & - & - & - \\ 
$Q1_{C_L} - C_L < 2.280$ & 194 & $247 \pm 41$ &$12 \pm 6$ & $0.4$ & $0.44 \pm 0.07$ & $13 \pm 6$ & $0.5$ \\ 
$Q2_{C_L} - 2.280 \leq C_L < 2.480 $ & 195 & $175 \pm 52$ &$4.4 \pm 3.9$ & $0.9$ & $0.29 \pm 0.10$ & $3.9 \pm 3.9$ & $1.2$ \\ 
$Q3_{C_L} - 2.480 \leq C_L < 2.652$ & 195 & $246 \pm 36$ &$12 \pm 5$ & $1.2$ & $0.43 \pm 0.06$ & $12 \pm 5$ & $1.6$ \\ 
$Q4_{C_L} - 2.662 \leq C_L < 2.862$ & 195 & $119 \pm 77$ &$1.4 \pm 2.7$ & $1.2$ & - & - & - \\
$\mathbf{Q5_{C_L} - C_L \geq 2.862}$ & \textbf{195} & $\mathbf{336 \pm 28}$ & $\mathbf{31 \pm 8}$ & $\mathbf{0.2}$ & $\mathbf{0.60 \pm 0.05}$ & $\mathbf{33 \pm 8}$ & $\mathbf{0.2}$ \\ 
$Q1_{C_L}$ to $Q4_{C_L}$ & 779 & $197 \pm 24$ &$6.2 \pm 2.3$ & $1.6$ & $0.34 \pm 0.04$ & $6.1 \pm 2.2$ & $2.5$ \\ 
$\mathbf{\Delta V < V_{lim}}$, $\mathbf{Q5_{C_L}}$ & \textbf{97} & $\mathbf{351 \pm 39}$ &$\mathbf{35 \pm 12}$ & $\mathbf{0.7}$ & $\mathbf{0.62 \pm 0.07}$ & $\mathbf{37 \pm 12}$ & $\mathbf{0.9}$ \\ 
$\mathbf{\Delta V < V_{lim}}$, $\mathbf{Q1_{C_L}}$ \textbf{to} $\mathbf{Q4_{C_L}}$ & \textbf{388} & $\mathbf{229 \pm 30}$ &$\mathbf{9.8 \pm 3.8}$ & $\mathbf{0.6}$ & $\mathbf{0.40 \pm 0.05}$ & $\mathbf{10.0 \pm 3.7}$ & $\mathbf{1.0}$ \\ 
$\mathbf{\Delta V > V_{lim}}$, $\mathbf{Q5_{C_L}}$ & \textbf{98} & $\mathbf{344 \pm 40}$ &$\mathbf{33 \pm 12}$ & $\mathbf{0.2}$ & $\mathbf{0.61 \pm 0.07}$ & $\mathbf{35 \pm 12}$ & $\mathbf{0.2}$ \\ 
$\Delta V > V_{lim}$, $Q1_{C_L}$ to $Q4_{C_L}$ & 390 & $141 \pm 47$ &$2.3 \pm 2.3$ & $1.0$ & $0.23 \pm 0.09$ & $1.9 \pm 2.2$ & $1.3$ \\ 
  
\hline         
\end{tabular}
\medskip
\begin{flushleft}
\textbf{Notes.} Columns: (1) Selection criteria; (2) number of groups considered in the stack; (3) and (4) results from the SIS profile fit, velocity dispersion and $M^{SIS}_{200}$; (5) and (6), results from the NFW profile fit, $R_{200}$ and $M^{NFW}_{200}$. Fields with ``-'' correspond to values that could not be determined through the lensing analysis given the low signal obtained. Well determined lensing masses 
(inconsistent with no signal at 2.5$\sigma$) are highlighted in bold.
\end{flushleft}
\label{table:results}
\end{table*}
\subsubsection{$\mu$ criterion}

\begin{figure*}
\includegraphics[scale=0.4]{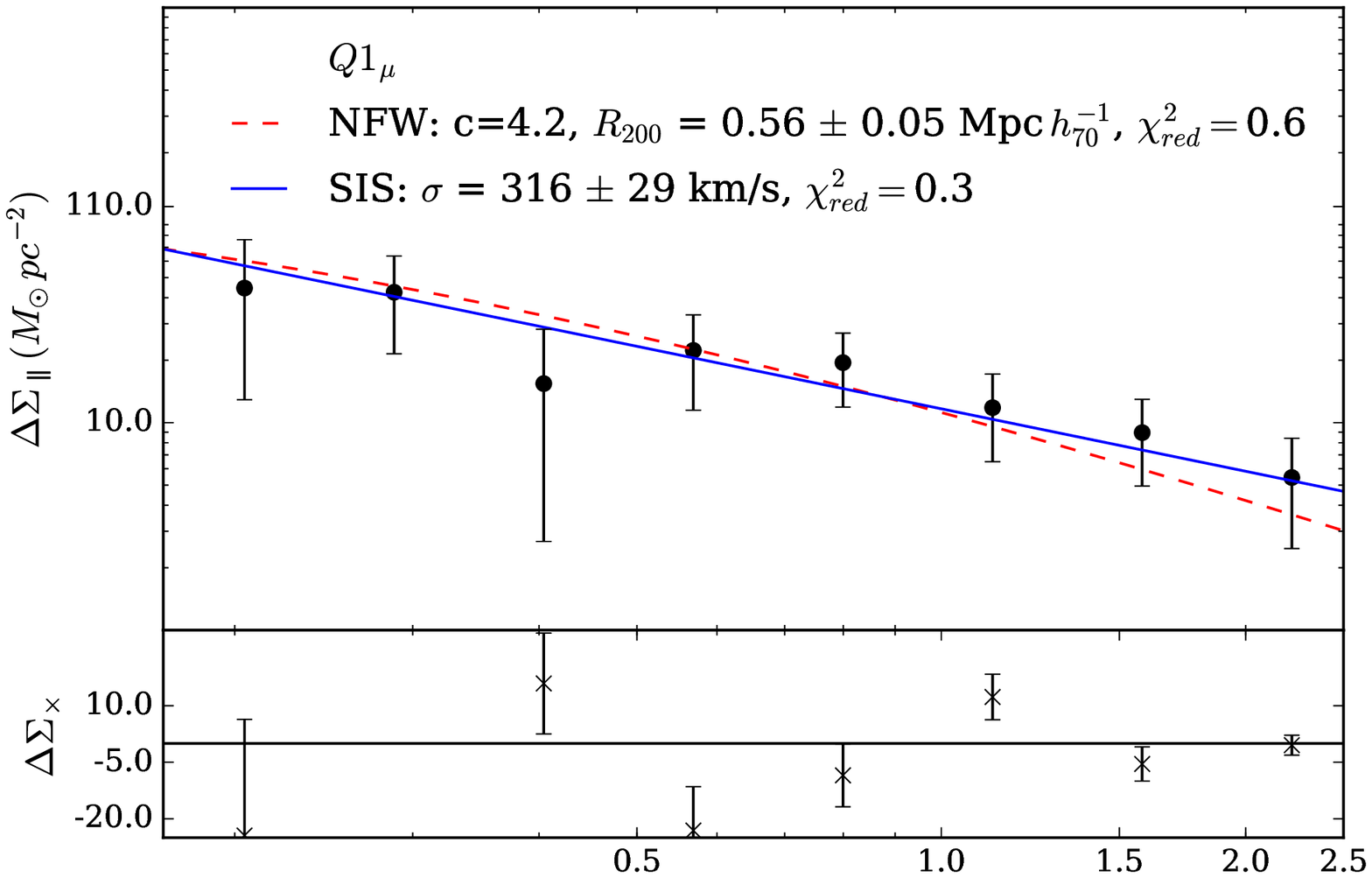}
\includegraphics[scale=0.4]{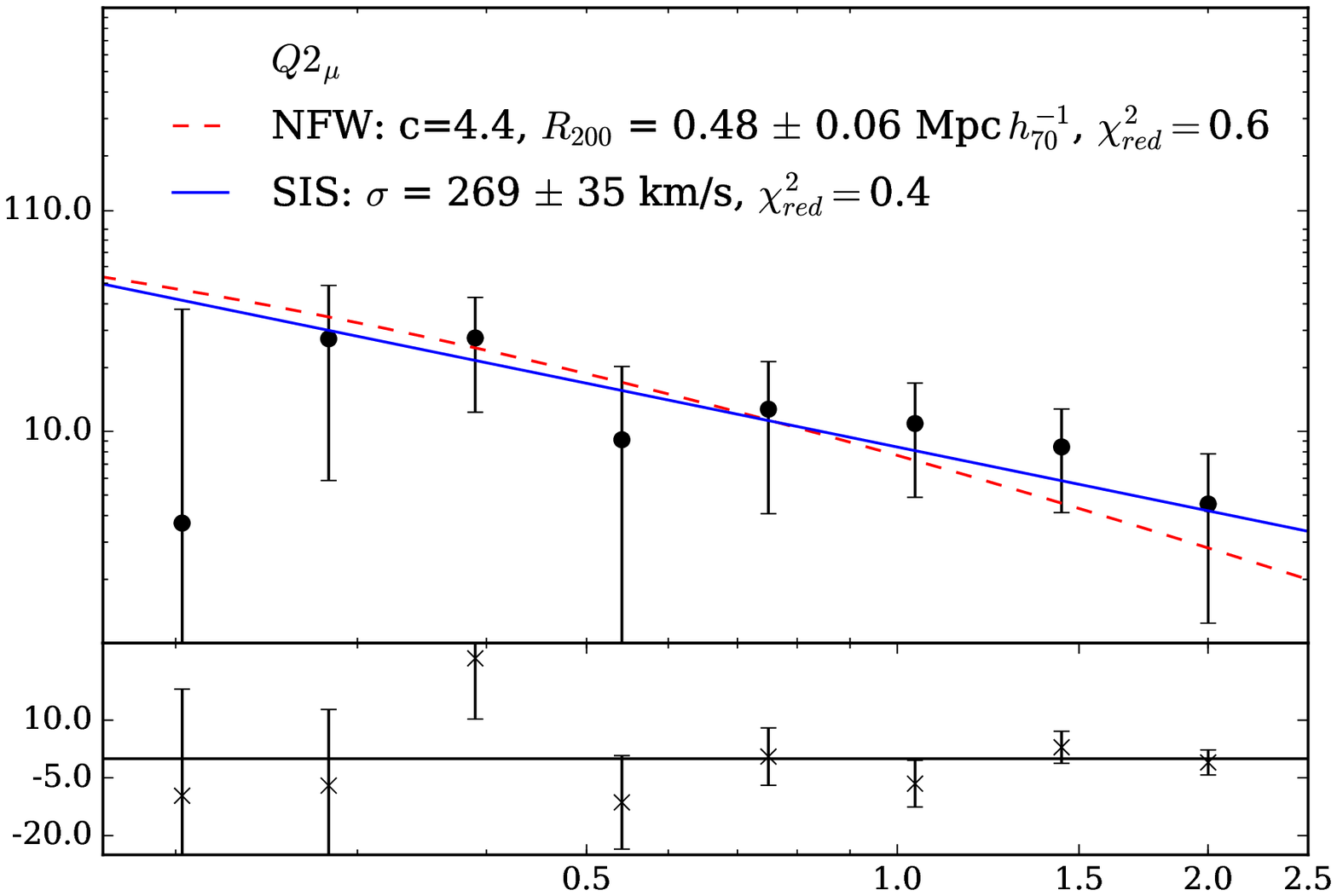}\\
\includegraphics[scale=0.4]{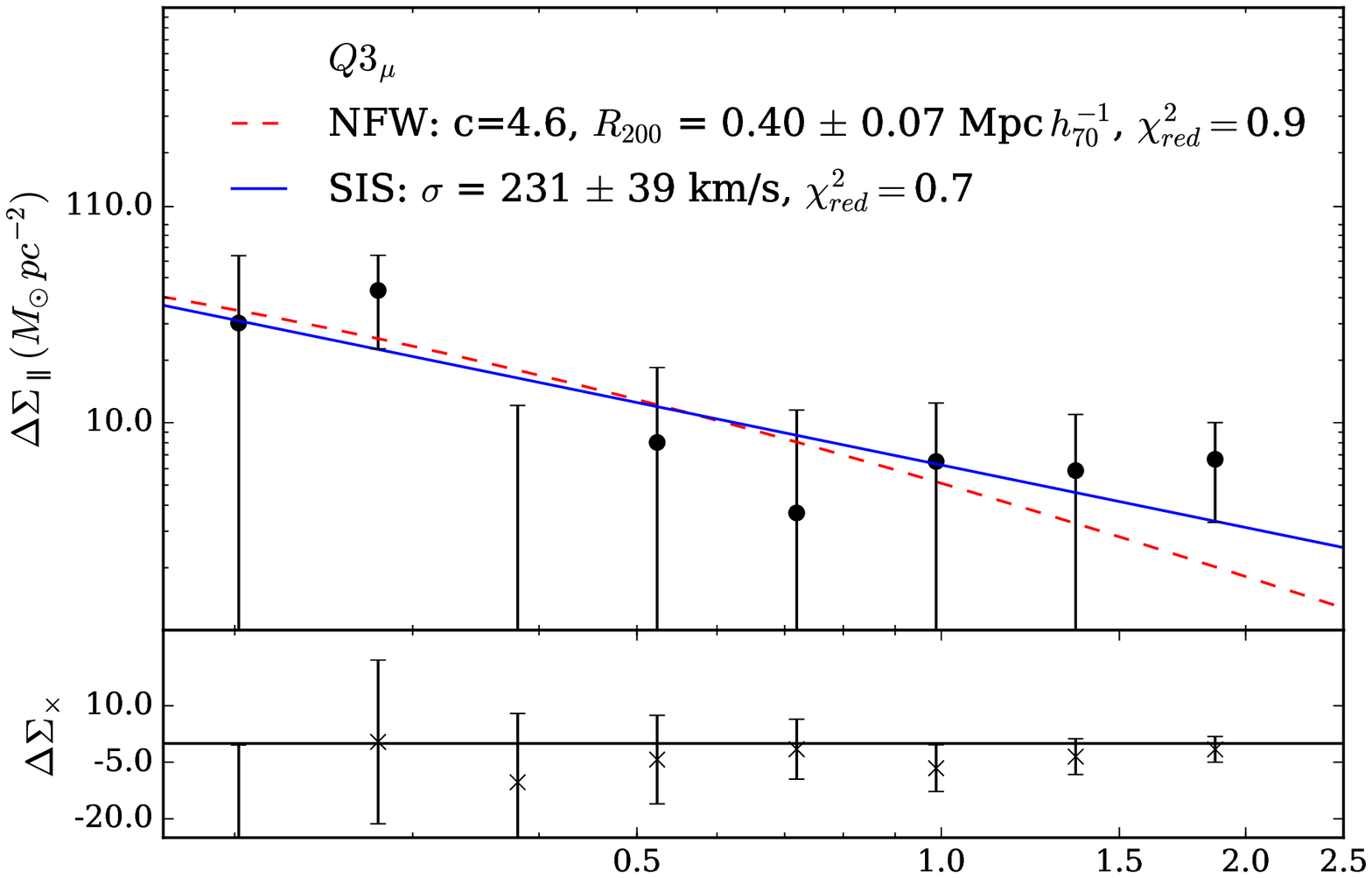}
\includegraphics[scale=0.4]{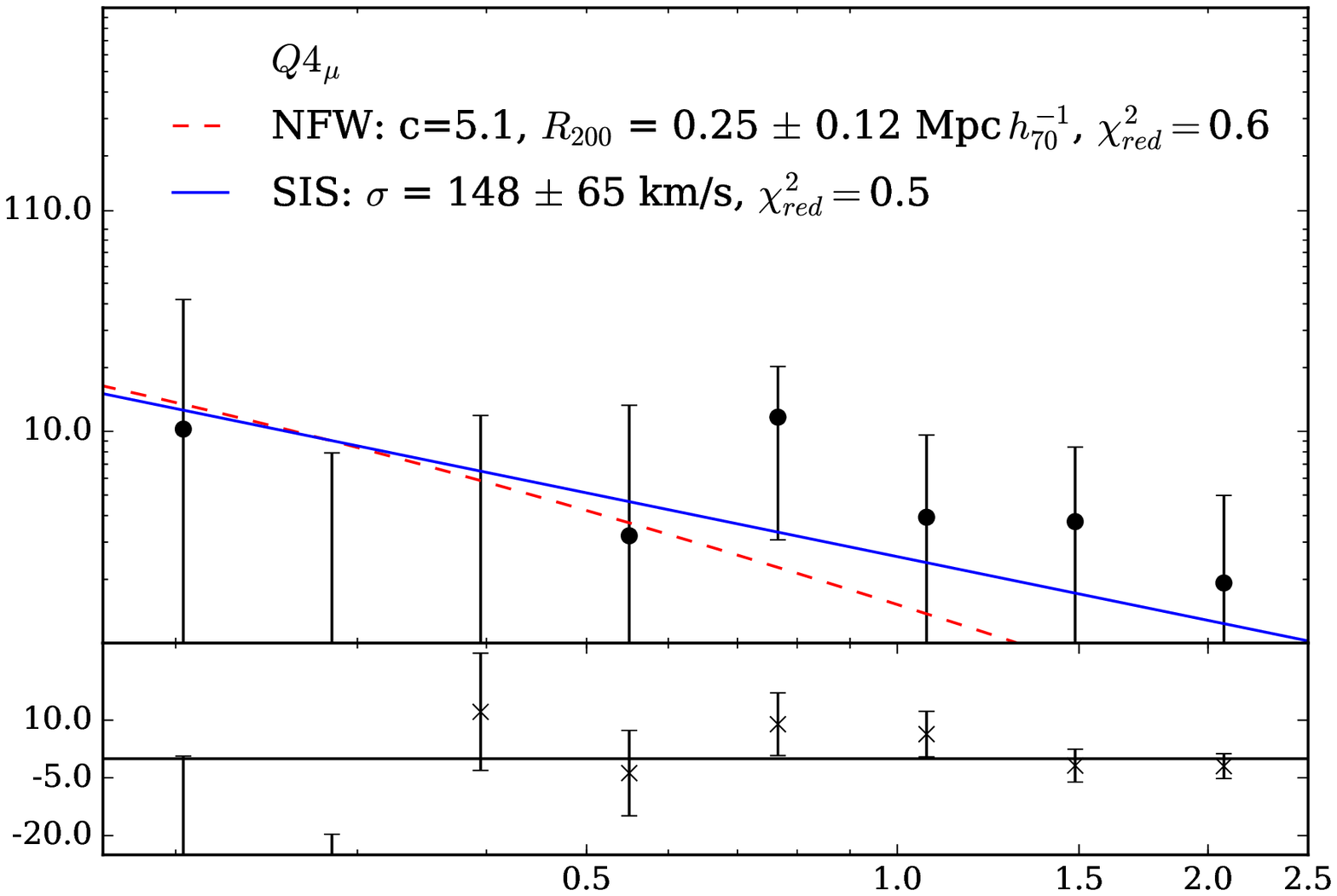}\\
\includegraphics[scale=0.4]{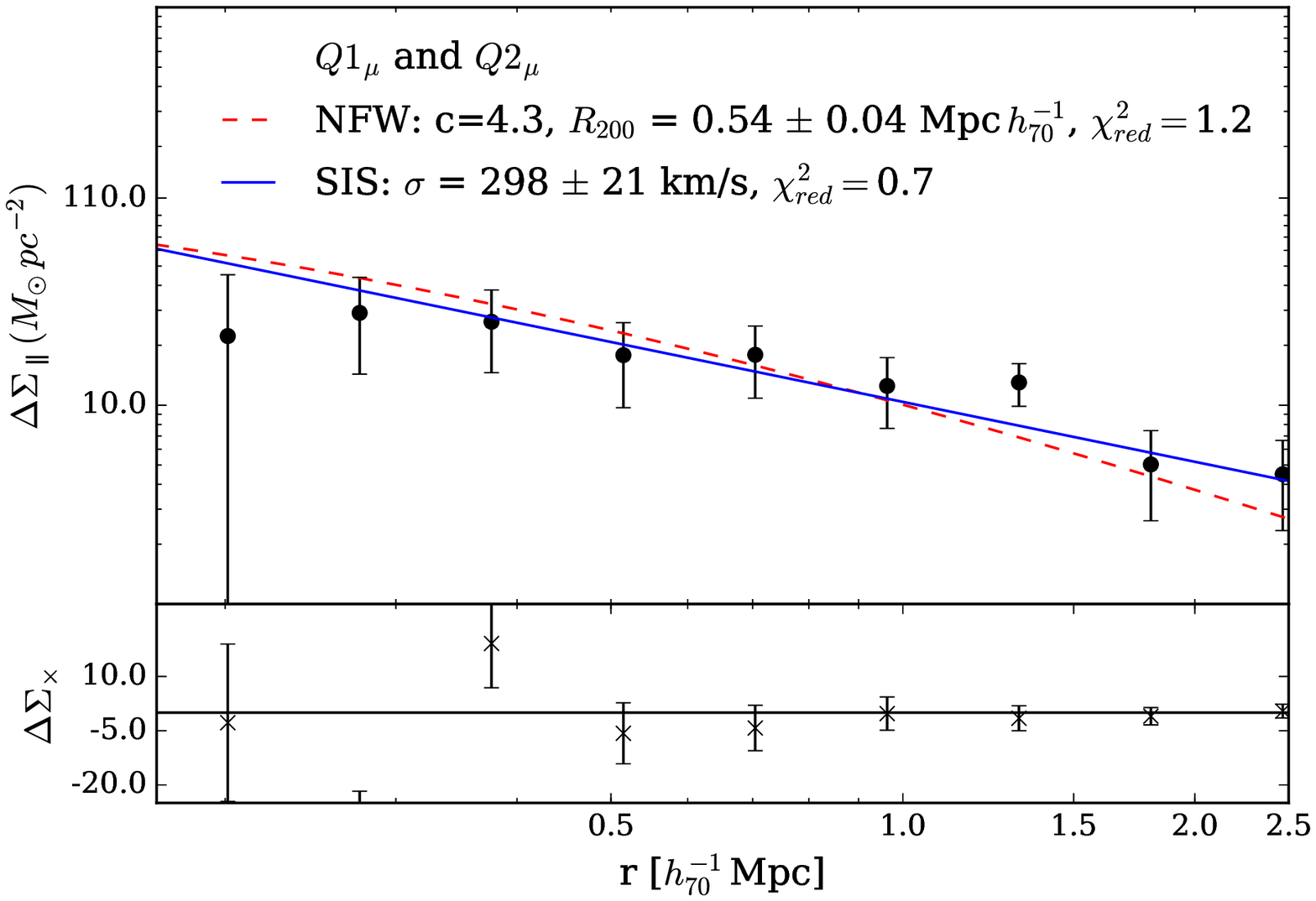}
\includegraphics[scale=0.4]{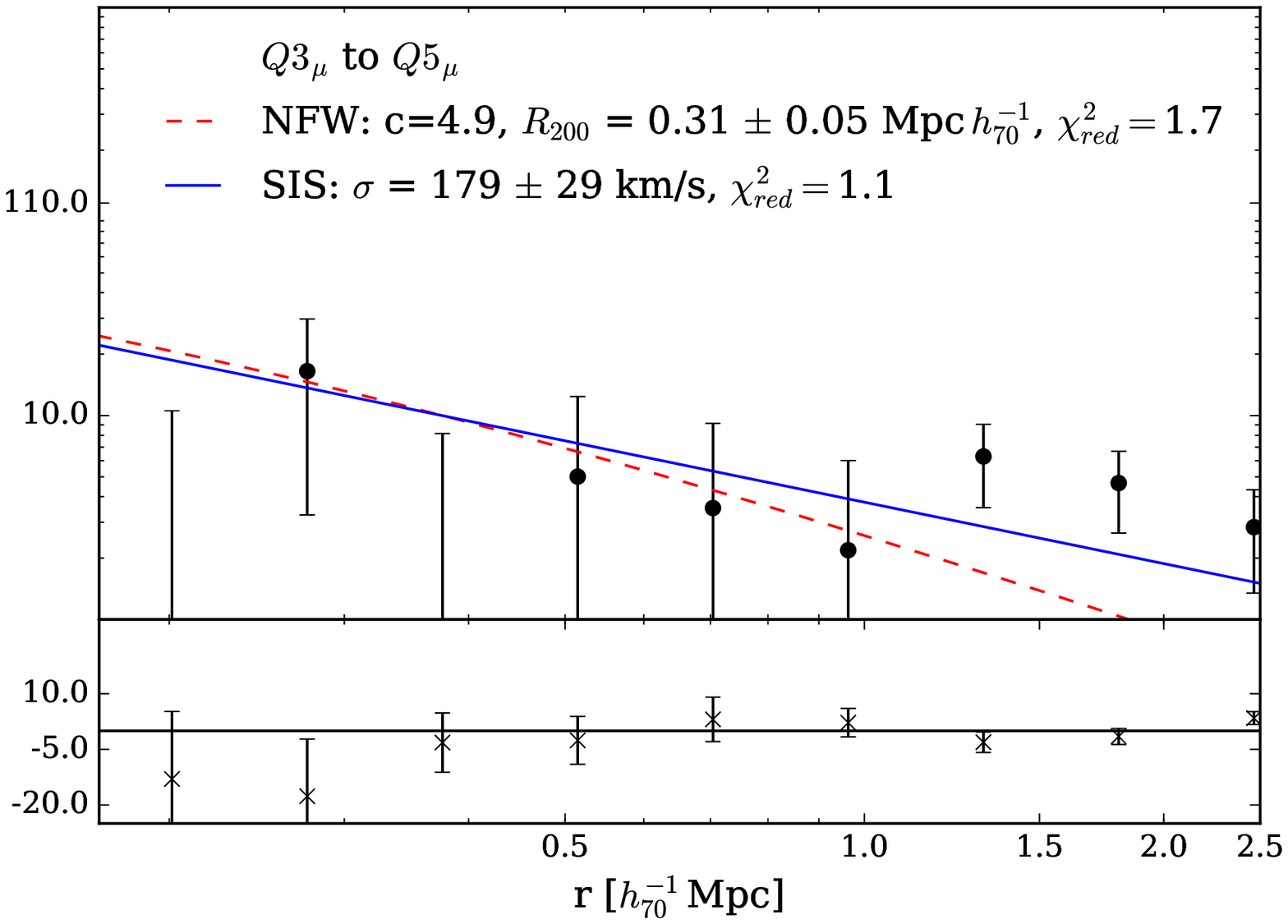}
\caption{Average density contrast profiles for the different $\mu$ quintiles.}
\label{fig:mu_quintil}
\end{figure*}

According to \citet{McConnachie2008} the number of CGs with interlopers decreases when a higher surface brightness cut is considered. To test how the lensing signal is affected by this parameter we divide the total sample into five equally-sized bins (i.e. quintiles), obtaining $\sim200$ groups per bin. Their resulting profiles are shown in Figure \ref{fig:mu_quintil}. The low $\mu$ quintile ($Q5_\mu$ subsample) is consistent with no signal and is not shown in the figure. As it can be seen, higher $\mu$ quintiles are consistent with a higher lensing signal. The quintiles subsamples with well determined lensing masses and adequately fitted profiles by the adopted models are $Q1_\mu$ and $Q2_\mu$. These subsamples together show a significant signal consistent with a SIS velocity dispersion of  $298 \pm 21$km\,s$^{-1}$ (Figure \ref{fig:mu_quintil}). On the other hand, the $Q3_\mu$, $Q4_\mu$ and $Q5_\mu$ subsamples together show a negligible signal in spite of the large number of systems.\\

\subsubsection{Concentration criterion}

\begin{figure*}
\includegraphics[scale=0.4]{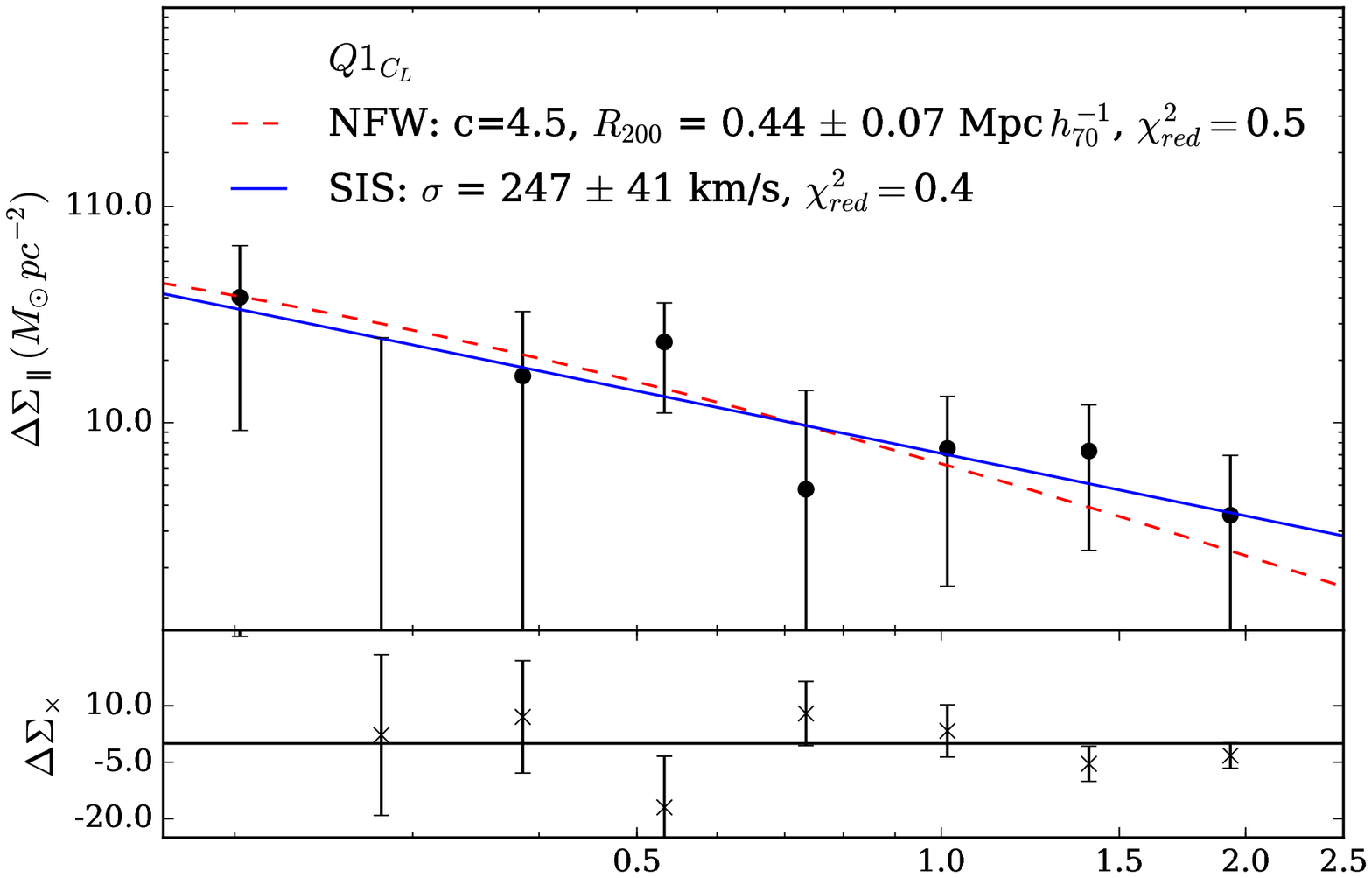}
\includegraphics[scale=0.4]{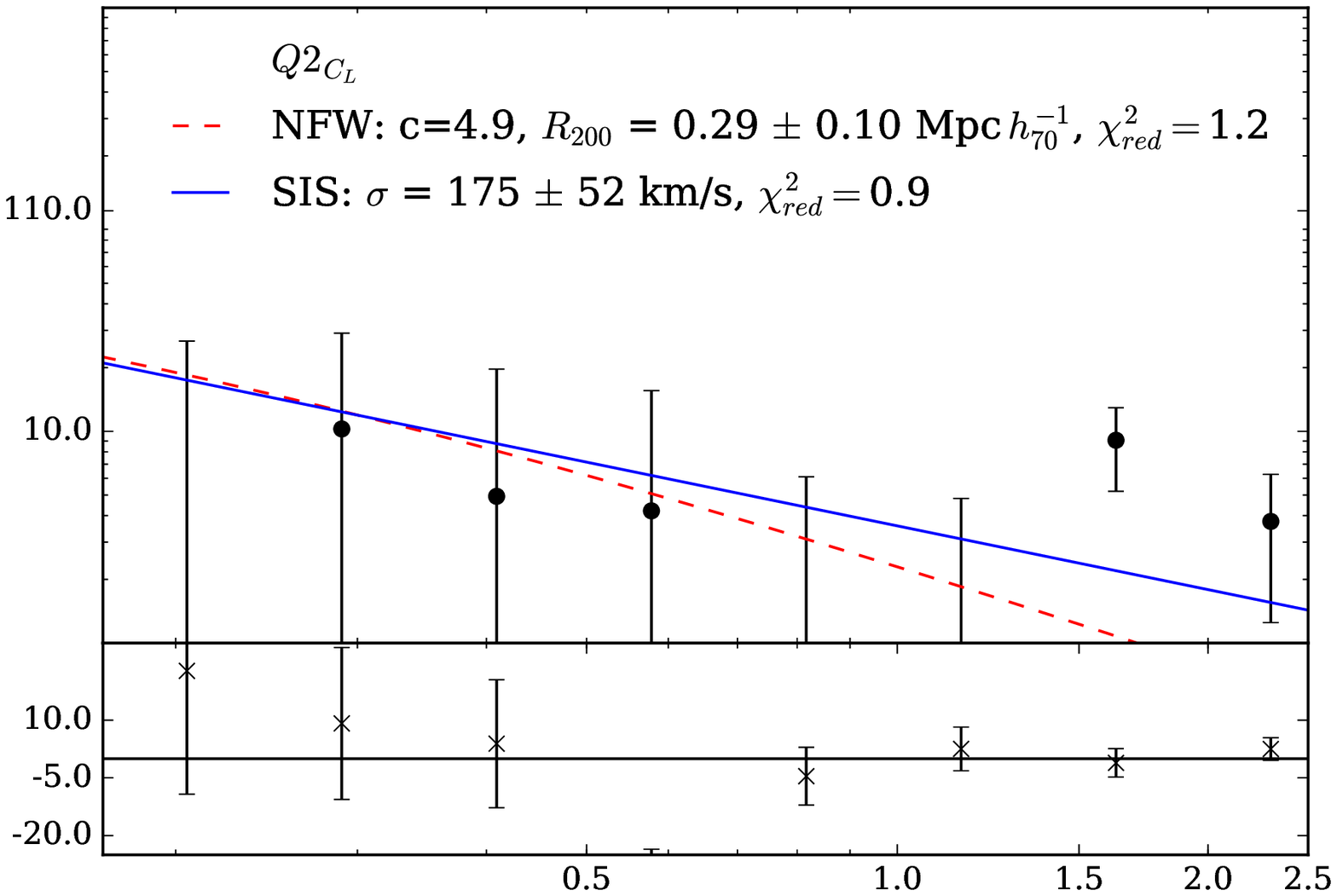}\\
\includegraphics[scale=0.4]{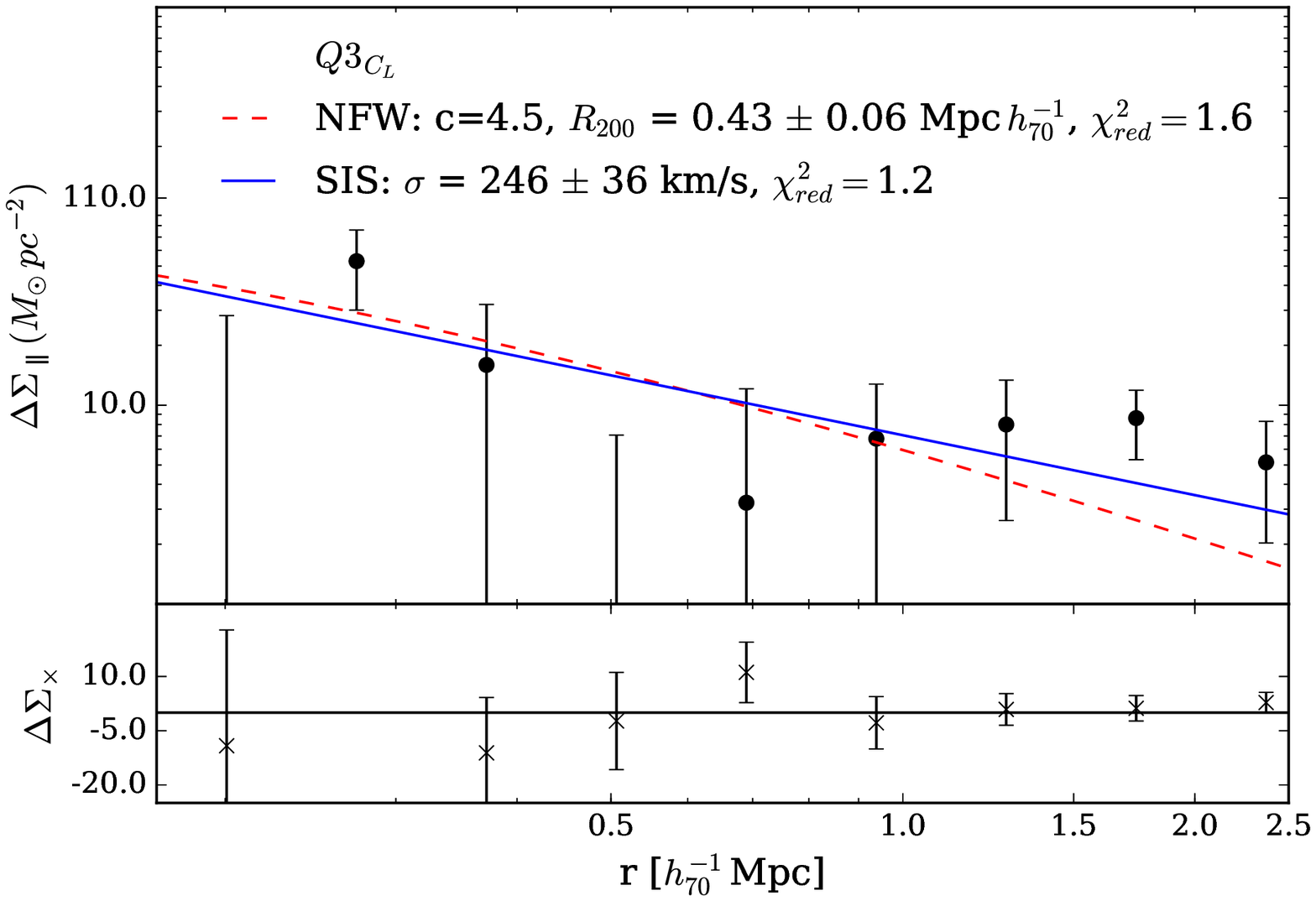}
\includegraphics[scale=0.4]{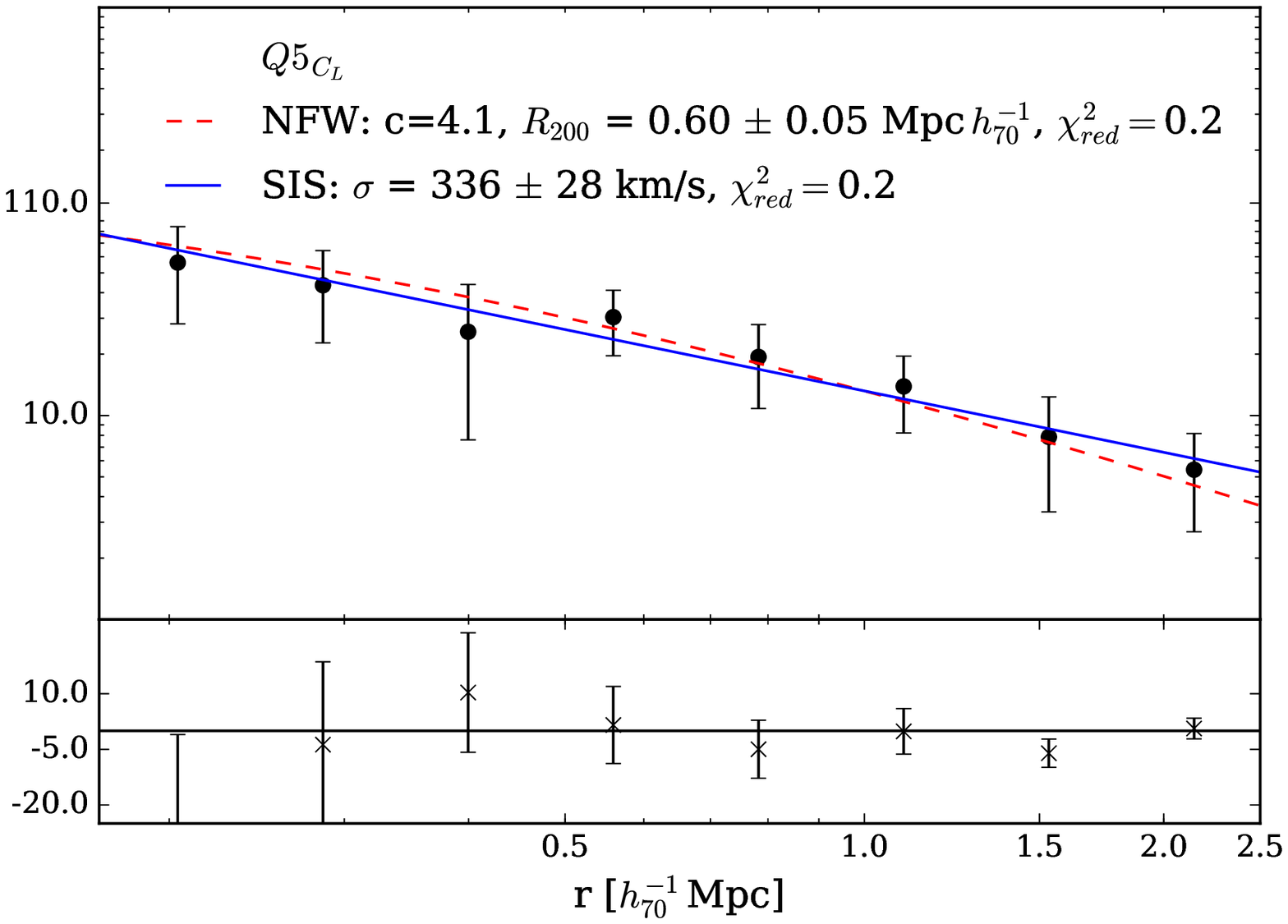}\\
\includegraphics[scale=0.4]{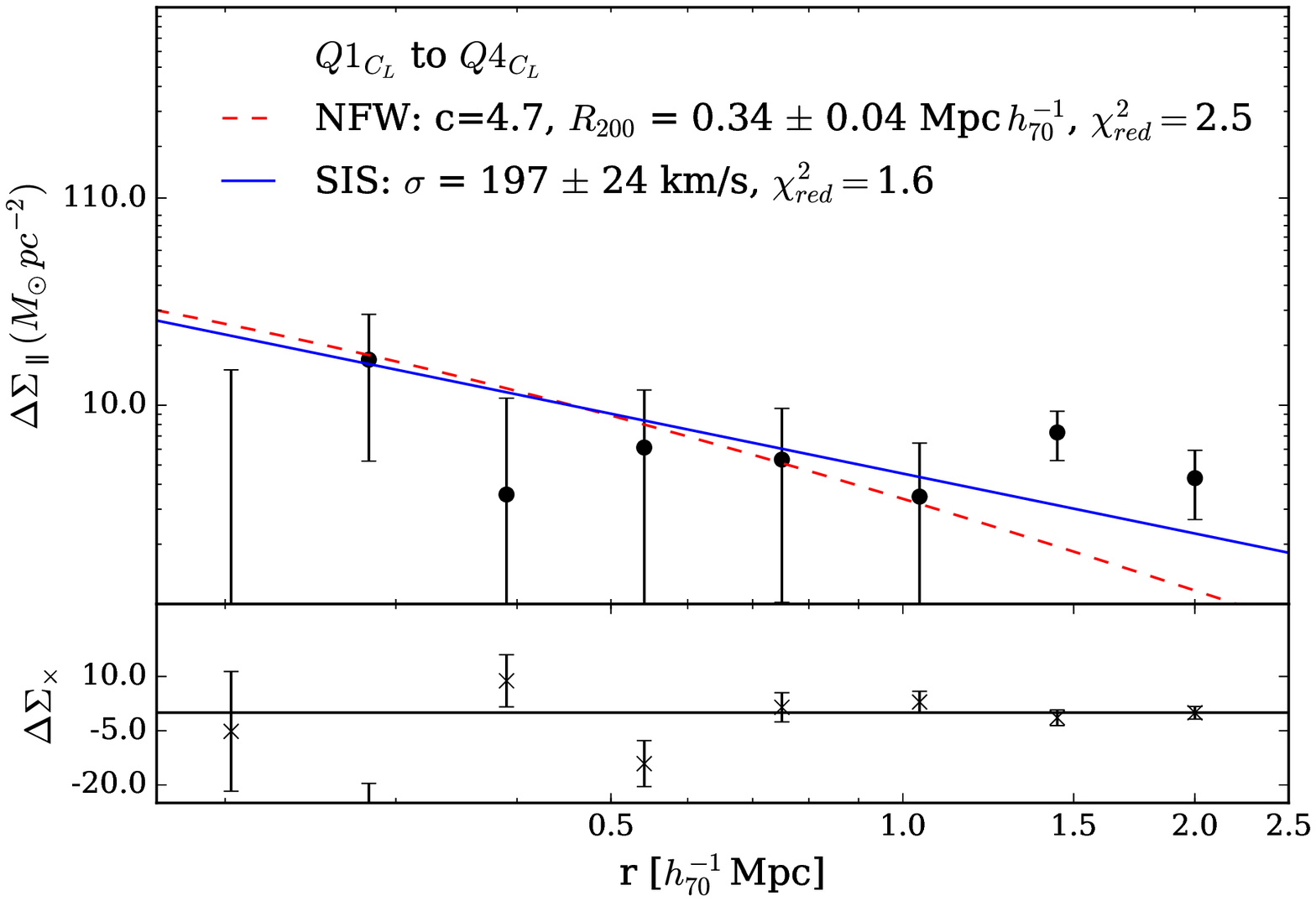}
\caption{Average density contrast profiles for the different $C_L$ quintiles. }
\label{fig:cw_quintil}
\end{figure*}

As in Paper\,I, we analyse the dependence of the lensing signal on the group average concentration index weighted by luminosity, $C_L$, defined as:
\begin{equation}
C_L = \frac{\sum c_i L_i}{\sum L_i},
\end{equation}
where sums run over the number of member galaxies, $L_i$ is the $r$-band luminosity and  $c_i$ is the galaxy concentration index defined as the ratio of the radii enclosing 90\% and 50\% of the Petrosian flux, i.e. $c_i = r_{90}/r_{50}$, obtained from the SDSS database. To compute the rest-frame luminosity $L_i$ of galaxies we $k$-correct magnitudes using \citet{Omill2011} code. \\
Following the analysis of surface brightness, we divide the sample into quintiles according to the $C_L$ distribution. As it can be seen in Figure \ref{fig:cw_quintil}, we obtain a significant lensing signal for the highest $C_L$ quintile $(\sigma_V = 336 \pm 28$ km\,s$^{-1}$) consistent with the results of Paper\,I. In contrast, the remaining quintiles with lower $C_L$ values have negligible lensing signal. The $Q4_{C_L}$ quintile is consistent with no signal and is not shown in the figure. Furthermore, the combination of these quintiles, $Q1_{C_L}$ to $Q4_{C_L}$, shows a slightly detectable signal, but it is not as well described by the adopted models as the $Q5_{C_L}$ profile (larger $\chi^2$ values).\\

\begin{figure}
\includegraphics[scale=0.5]{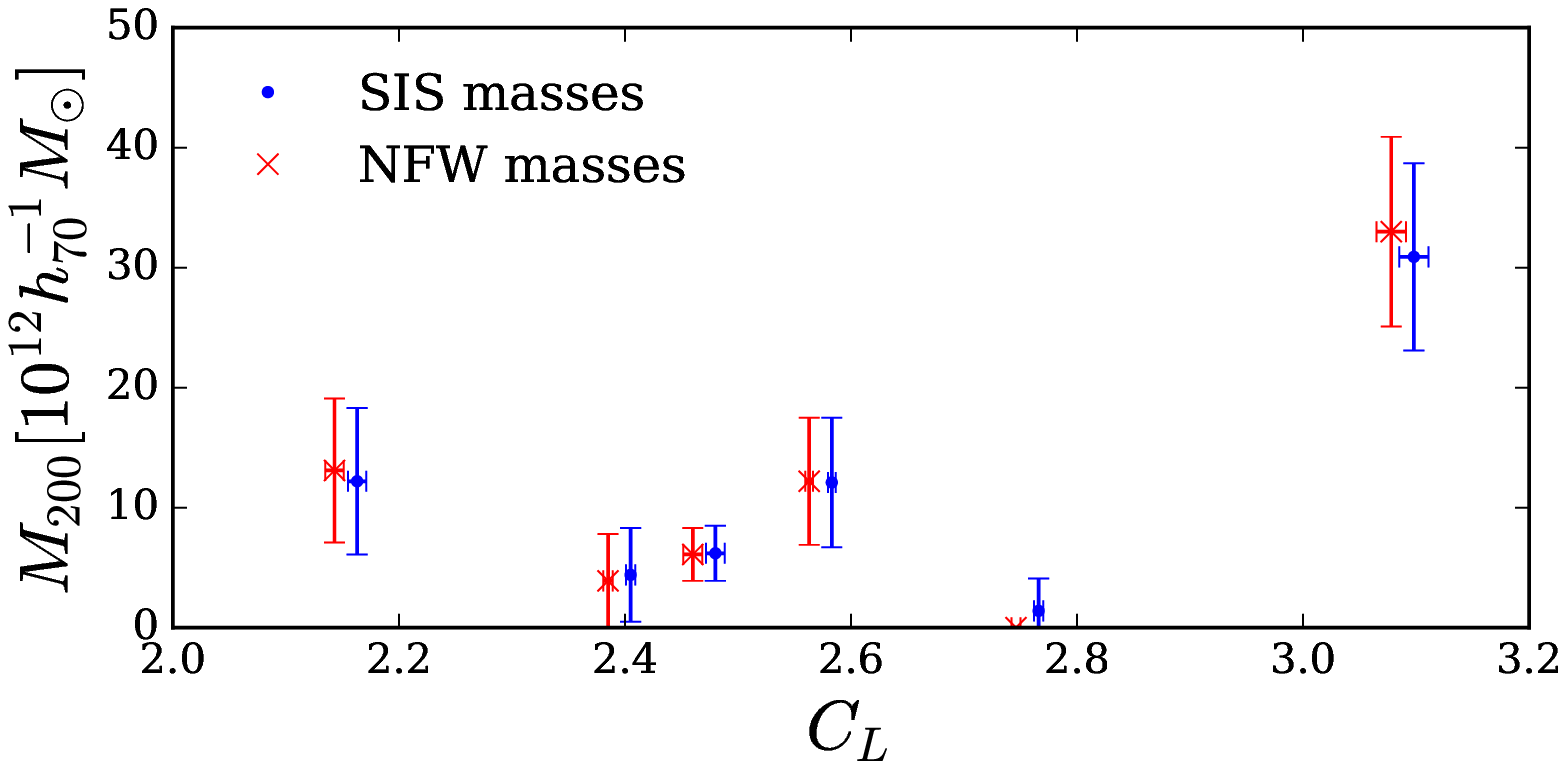}
\includegraphics[scale=0.5]{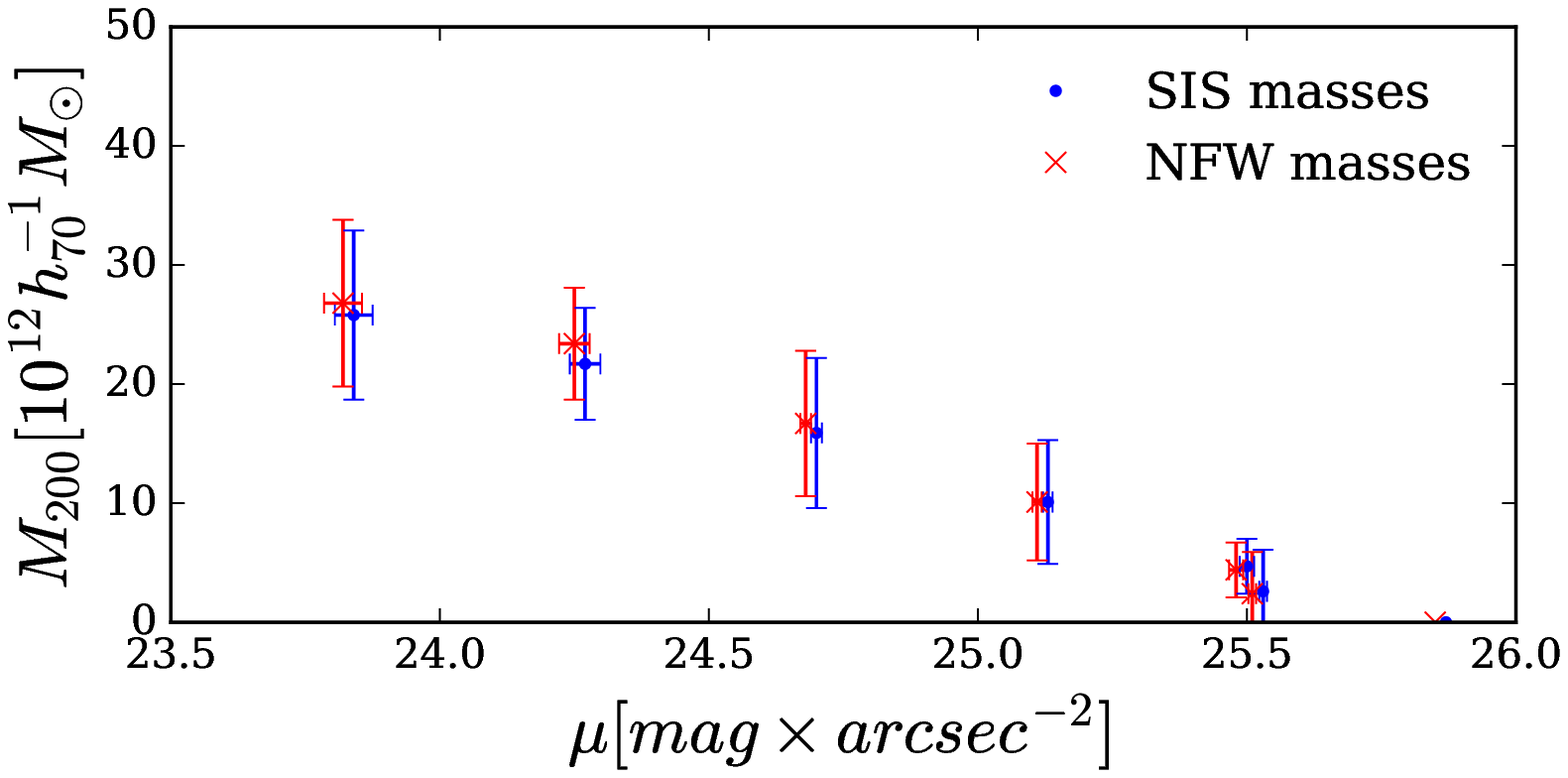}
\caption{Obtained masses for the average $\mu$ and $C_L$ subsamples considered. Values for the SIS model are slightly shifted for more clarity.}
\label{fig:param_dependence}
\end{figure}
\subsubsection{Analysis of the adopted criteria}

\begin{figure}
\includegraphics[scale=0.4]{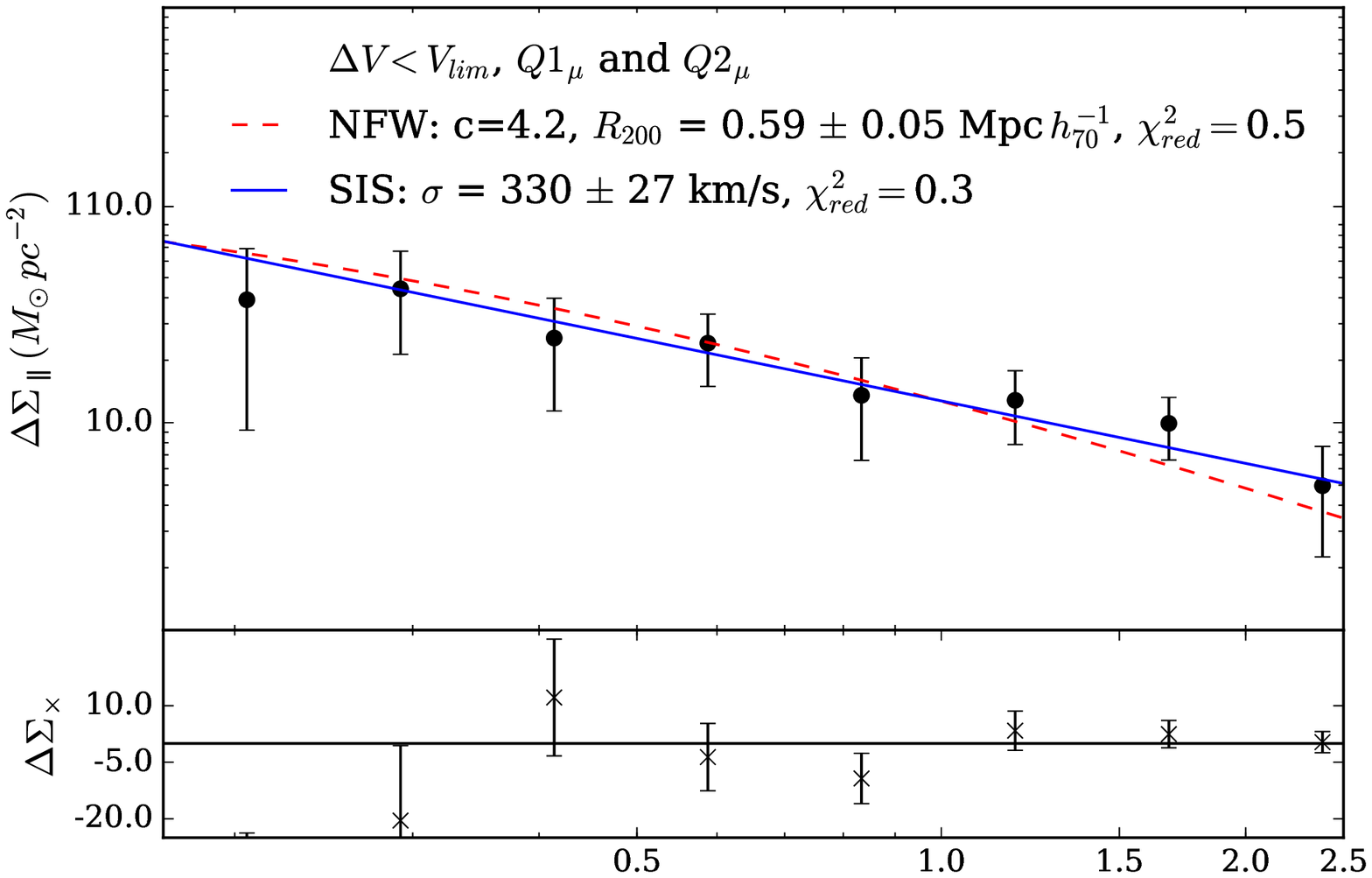}\\
\includegraphics[scale=0.4]{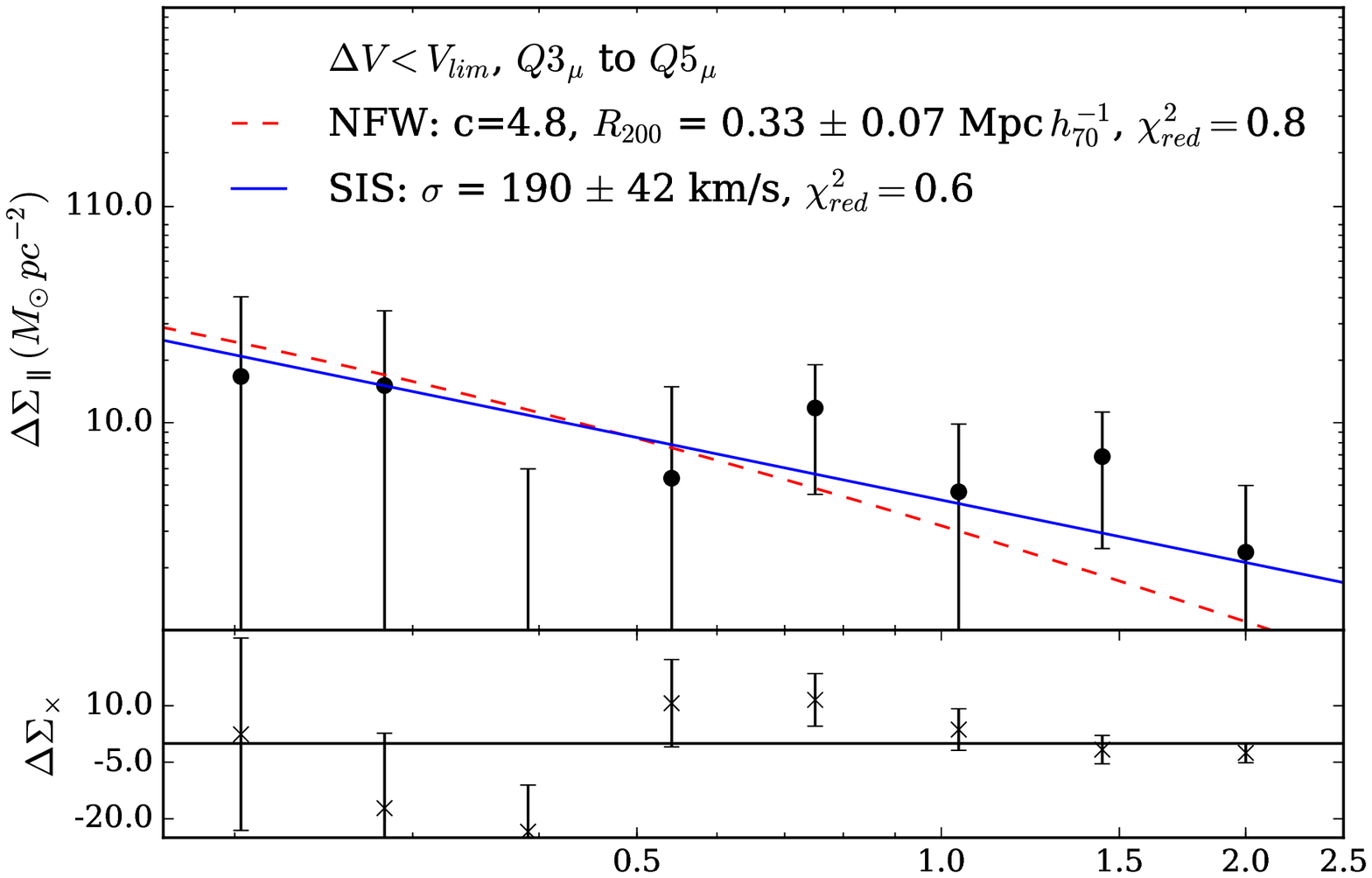}\\
\includegraphics[scale=0.4]{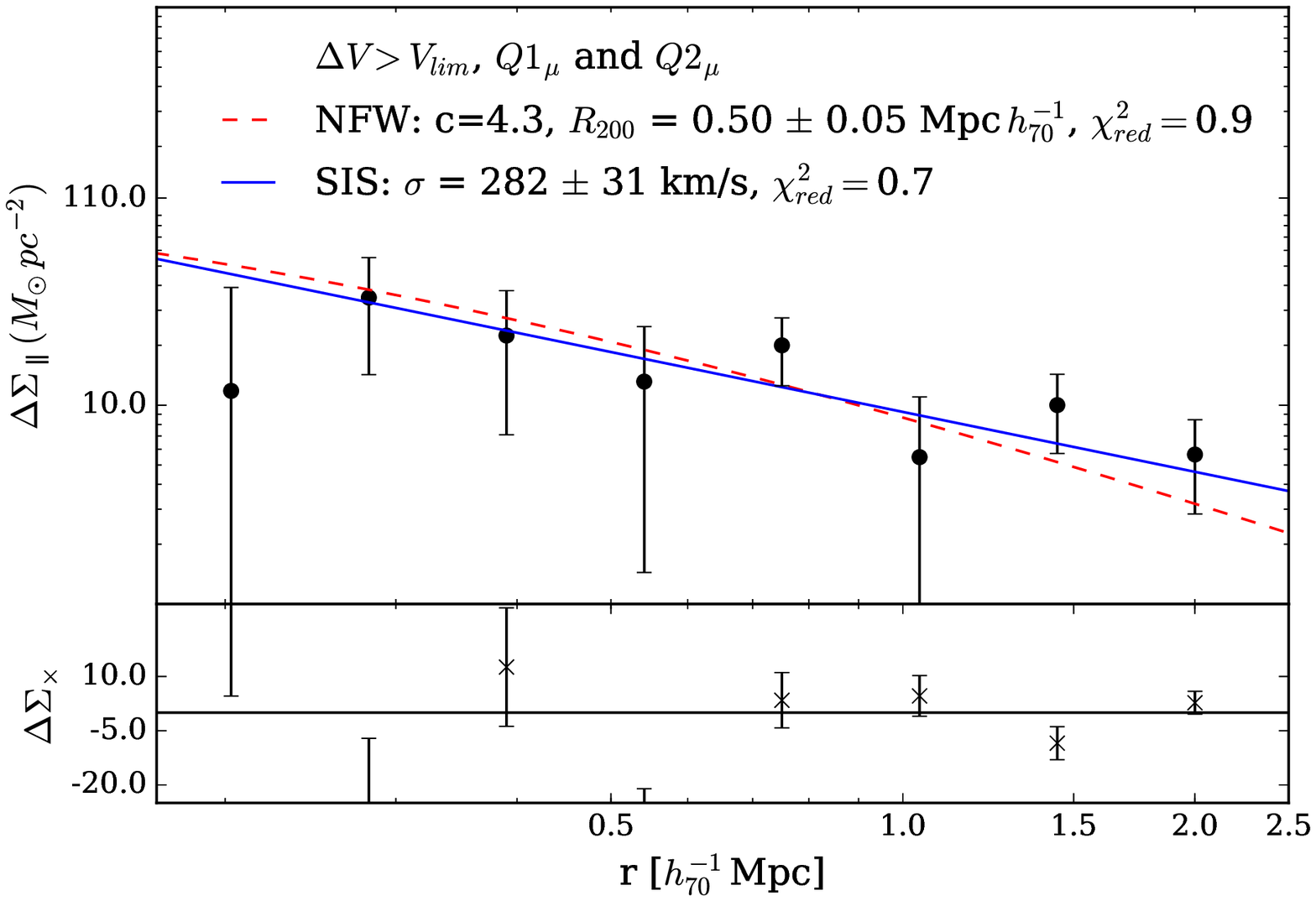}
\caption{Average density contrast profiles for $\Delta V$ and $\mu$ criteria. From top to bottom: higher $\mu$ half of $\Delta V < V_{lim}$ subsample, lower $\mu$ half of $\Delta V < V_{lim}$ subsample, higher $\mu$ half of $\Delta V > V_{lim}$ subsample.}
\label{fig:mu_halves}
\end{figure}

\begin{figure}
\includegraphics[scale=0.4]{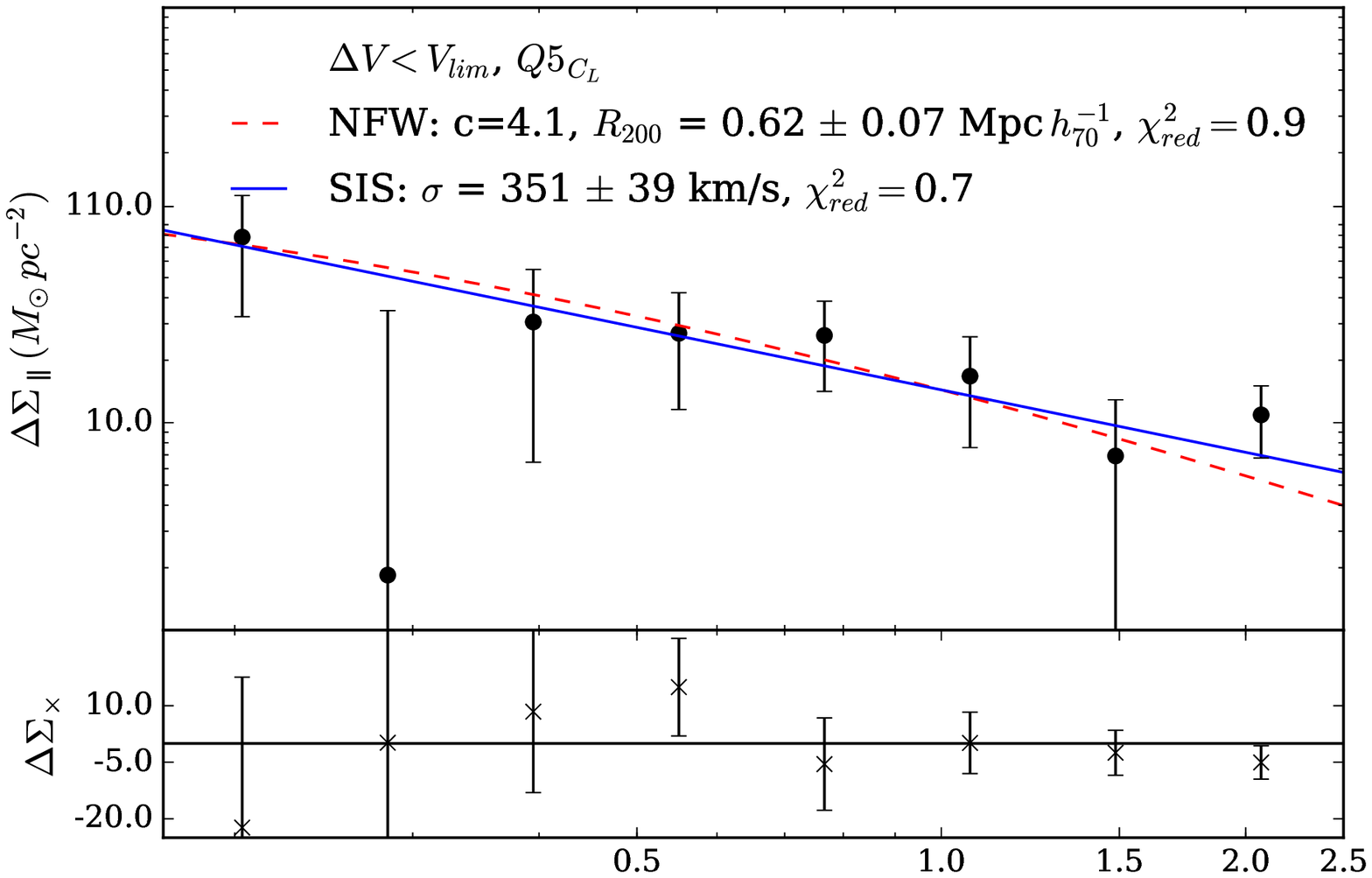}\\
\includegraphics[scale=0.4]{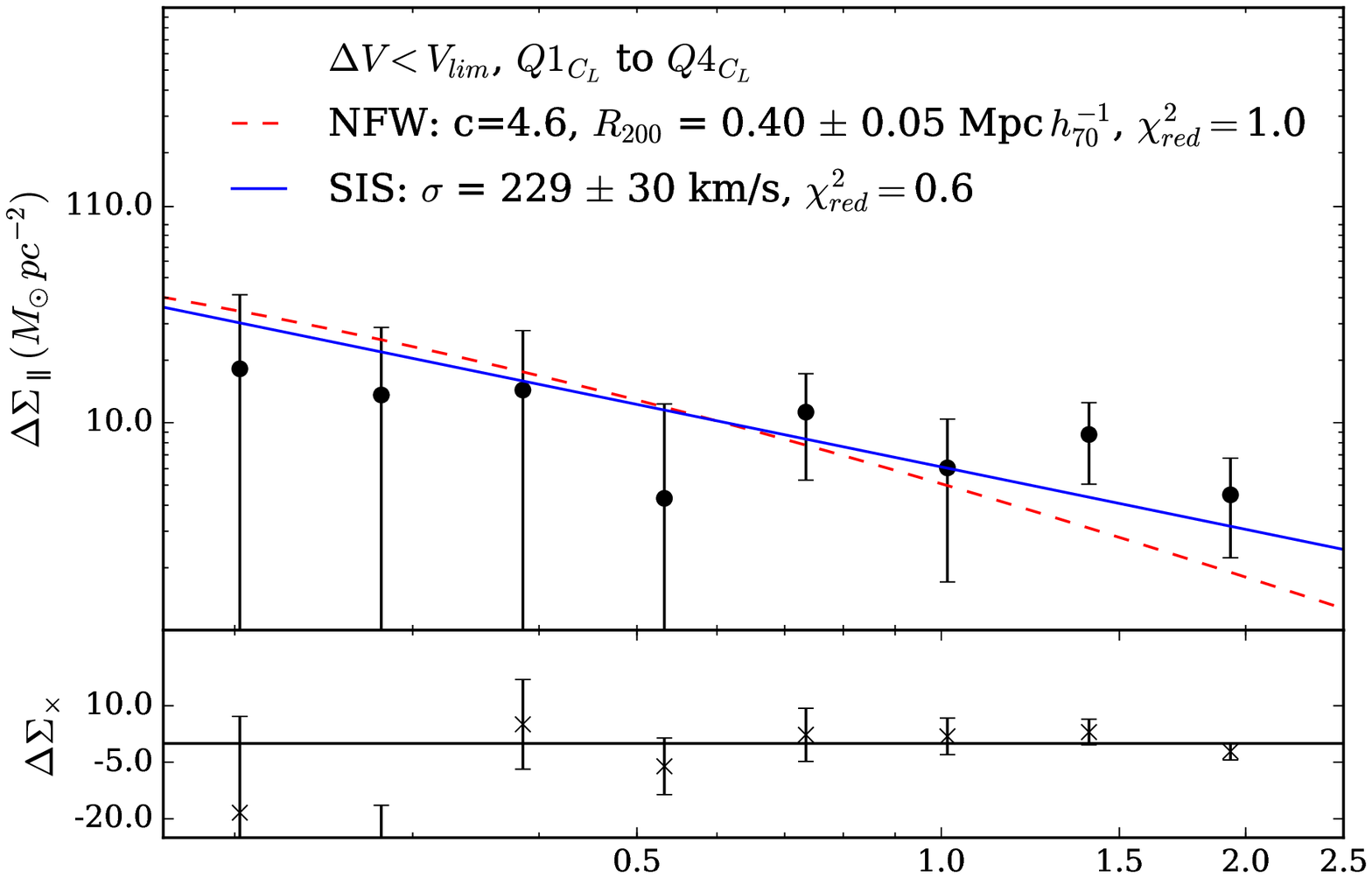}\\
\includegraphics[scale=0.4]{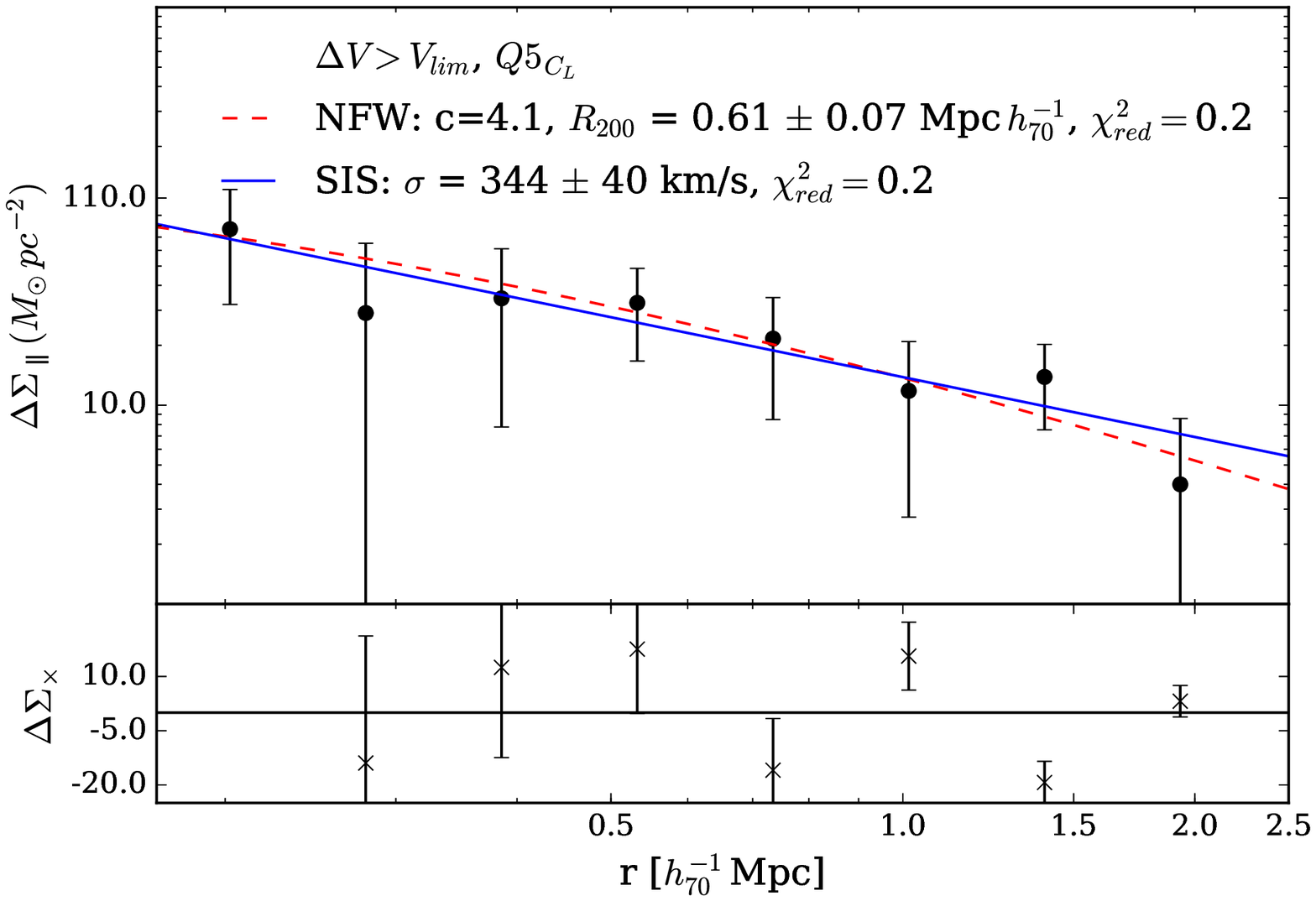}
\caption{Average density contrast profiles for $\Delta V$ and $C_L$ criteria. From top to bottom: lower $C_L$ half of $\Delta V < V_{lim}$ subsample, higher $C_L$ half of $\Delta V < V_{lim}$ subsample, higher $C_L$ half of $\Delta V > V_{lim}$ subsample. }
\label{fig:cw_halves}
\end{figure}

As shown in this section, the lensing signal and the $C_L$ and $\mu$ parameters are correlated. In Figure \ref{fig:param_dependence} we show the lensing masses vs. these parameters. It can be noticed that high mass values are obtained only in the highest $C_L$ quintile ($Q5_{C_L}$ subsample), contrary to the gentle increase in mass values with surface brightness. Also, it can be seen that the NFW and SIS masses are in mutual agreement for all the subsamples considered.\\
We also divide the $\Delta V < V_{lim}$ and $\Delta V > V_{lim}$ subsamples into quintiles in, $\mu$ and $C_L$. The corresponding profiles are shown in Figures \ref{fig:mu_halves} and \ref{fig:cw_halves}. We do not obtain a significant lensing signal for low surface brightness and concentration index when considering CGs with $\Delta V > V_{lim}$. On the other hand, we find that the $\Delta V$ cut has no impact on the lensing signal of groups with larger surface brightness and $C_L$ values, which would tend to include a larger fraction of gravitationally bound systems.

\subsection{Systematic errors and control test}
\label{subsec:systematics}
We analyse the relation between the lensing signal with different photometric redshift cuts to select background galaxies, in order to verify if there is any bias introduced by the adopted criteria (ODDS\_BEST $>$ 0.6 and $z > z_g+0.1$). Following \citet{Leauthaud2017} analysis, we compute the lensing profile for different ODDS\_BEST cuts (ODDS\_BEST $>$ 0, ODDS\_BEST $>$ 0.2, ODDS\_BEST $>$ 0.4, ODDS\_BEST $>$ 0.6 and ODDS\_BEST $>$ 0.7) for the samples with better quality signal ($\Delta V < V_{lim}$, $Q1_{\mu}$ and $Q2_{\mu}$, and $Q5_{C_L}$ samples) and for the total sample. We observe an increase in the lensing signal when we restrict the source galaxy sample to larger ODDS\_BEST cuts. This could be due to a lower contamination by group members and foreground galaxies as a result of higher quality redshift estimates. Nevertheless, derived mass profile parameters are in agreement within $\sim 1.5\sigma$. We also observe an increase of the parameter errors ($\sim 40\%$) from  ODDS\_BEST $>$ 0.6 to ODDS\_BEST $>$ 0.7 due to the low number of background galaxies. Therefore we conclude that a ODDS\_BEST $>$ 0.6 cut is a suitable compromise between quality and number of background galaxies. We have also tested if the $z > z_g+0.1$ criteria is sufficient considering the large uncertainties in the photometric redshifts. Therefore, we compute the profile with different lens-source separation cuts ($z > z_g+0.1$, $z > z_g+0.2$ and $z > z_g+0.3$).  No statistically significant systematic trend is found in this test, even in the inner regions where the contamination by group member galaxies could be significant and dilute the lensing signal. Also, derived fitted parameters are consistent within $0.3\sigma$.\\%
Finally, we perform a nullity control test assigning to each group a random centre and select background galaxies in the same angular range as for the groups. The resultant profiles are consistent with a null signal for the tangential and the cross component.

\subsection{Comparison with dynamical estimates}
\label{subsec:dynamics}
\begin{figure}
\includegraphics[scale=0.4]{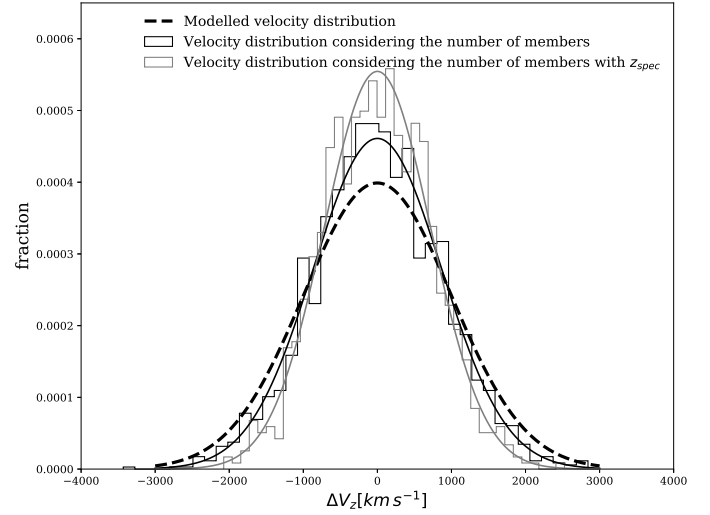}
\caption{Normalized line-of-sight velocity distributions. Dashed line corresponds to a modelled Gaussian of 1000\,km\,s$^{-1}$. Black histogram corresponds to the distribution obtained by taking random velocities considering the number of members in the CGs sample; the black curve is the gaussian fit of this distribution, wich has a dispersion of 877\,km\,s$^{-1}$. Grey histogram corresponds to the distribution obtained by taking random velocities of the modelled Gaussian, considering the number of members with spectroscopic measurements in the CGs sample with more than two $z_{spec}$; the gray curve is the gaussian fit of this distribution, which has a dispersion of 710\,km\,s$^{-1}$.}
\label{fig:dv_dist}
\end{figure}

\begin{table}

\caption{Comparison with dynamical velocity dispersion estimates.}
\begin{tabular}{@{}cccc@{}}
\hline
\hline
\rule{0pt}{1.05em}%
  Selection criteria  &$\sigma_{V}$ &  $\sigma_{dyn}$ &\\
   $\Delta V < V_{lim}$      & [km\,s$^{-1}$]    &   [km\,s$^{-1}$]  \\

 \hline
\rule{0pt}{1.05em}%
Total sample   				&  $266 \pm 22$  &  $307 \pm 19$	\\ 
$Q1_{\mu}$ and $Q2_{\mu} $  &  $330 \pm 27$  &  $305 \pm 22$	\\ 
$Q3_{\mu}$ to $Q5_{\mu} $   &  $190 \pm 42$  &  $309 \pm 27$	\\ 
$Q5_{C_L} $   				&  $351 \pm 39$  &  $322 \pm 24$	\\ 
$Q1_{C_L}$ to $Q4_{C_L}$	&  $229 \pm 30$  &  $302 \pm 19$	\\ 
\hline         
\end{tabular}
\medskip
\begin{flushleft}
\textbf{Notes.} Columns: (1) Selection criteria; (2) lensing velocity dispersion; (3) dynamical velocity dispersion (errors are computed considering bootstrap realizations and the dispersion in the correction factor).
\end{flushleft}
\label{table:results-deltaV}
\end{table}
We compare the results obtained for the samples with $\Delta V < V_{lim}$ with the dynamical values obtained by computing the mean galaxy velocity dispersion using the complete \citet{McConnachie2009} catalogue. To compute dynamical values we only consider groups with $\Delta V < 1000 \kms$ and with at least two spectroscopic estimates. Given the low number of tracers in each group to estimate the dynamical velocity dispersion (in the majority of cases 2 or 3 galaxies), we use a Monte-Carlo simulation to correct for this low-number statistic bias. We assume a fiducial Gaussian velocity distribution and simulate individual CGs with a given number of members with measured $z$ (this number of members with $z$ is randomly generated with the same distribution of real CGs with known $z$). Then, for each CG we compute the $\Delta V$ values and proceed to stack the distribution of all CGs. The velocity dispersion is computed as the standard deviation of this stacked $\Delta V$ distribution. In Figure \ref{fig:dv_dist} we show $\Delta V_z$ distributions for one realization, together with the modelled distributions. We remark that even in the scenario of 100\% redshift completeness, the velocity dispersion must be significantly corrected due to the low-number statistic bias. For the group sample considered we obtain that the measured dynamical dispersion is underestimated by $\sim 29 \%$, therefore a correction factor of $1.40 \pm 0.05$ is applied. The derived dynamical values are also shown in Table \ref{table:results-deltaV} together with the lensing estimates.\\
There is a good agreement (within $1\sigma$) between dynamical and lensing results for the total, high surface brightness and high concentration subsamples with $\Delta V < V_{lim}$. Interlopers bias the dynamical velocity dispersion to larger values, although, the lensing signal is weakened. Therefore, dynamical estimates are higher for the samples with larger contamination by interlopers ($Q3_{\mu}$ to $Q5_{\mu} $ and $Q1_{C_L}$ to $Q4_{C_L}$ subsamples).

\section{Conclusions}
\label{sec:conclusions}
In this work we present a lensing analysis of CGs using the CS82 survey. The results obtained are in agreement with those found in Paper\,I, confirming the previous lensing analysis using SDSS images, with lower depth and image quality than in the CS82 survey. We extend our previous analysis to the redshift range $0.2<z<0.4$ including groups with lower surface brightness. The number of members with redshift determinations was extended by using photometric redshifts.\\ 
We analyse the dependence of the lensing signal on the group surface brightness, the group concentration parameter and a limiting relative velocity criterion of members according to redshift information. A considerable improvement of the density profile is obtained for a group sample with members restricted to relative velocity $\Delta V < V_{lim}$. We also observe a correlation between the derived masses and $\mu$ and $C_L$ values. We obtain larger lensing masses for CGs with higher surface brightness. Similarly, CGs with more concentrated member galaxies tend to have larger lensing masses in agreement with Paper\,I. The largest lensing signal consistent with a SIS $\sigma \sim 336 \kms$ is obtained for groups with larger $C_L$ values ($Q5_{C_L}$ subsample). This subsample includes CGs with red, concentrated member galaxies, consistent with those observed in dense environments such as galaxy groups and clusters.\\
We compare the lensing results obtained for groups that satisfy $\Delta V < V_{lim}$ with dynamical estimates. There is a good agreement between derived dynamical dispersions and the lensing values for the total, high surface brightness and high concentration samples with $\Delta V < V_{lim}$. On the other hand, for the subsamples with larger contamination by interlopers, derived dynamical values are larger than the lensing results. An interpretation is that interlopers bias the dynamical velocity dispersion high, whereas the lensing signal is biased low. \\
In future works we will study the mass and mass-to-light ratio of CGs and their dependence on environment and redshift aiming to shed light on the formation and evolution of these systems.  
\section*{Acknowledgments}
The authors thank the anonymous referee for its useful comments
that helped improved this paper.\\
This work was partially supported by the Consejo Nacional de Investigaciones Cient\'{\i}ficas y T\'ecnicas (CONICET, Argentina) 
and the Secretar\'{\i}a de Ciencia y Tecnolog\'{\i}a de la Universidad Nacional de C\'ordoba (SeCyT-UNC, Argentina).\\
MM is partially supported by CNPq (grant 312353/2015-4) and FAPERJ. FORA TEMER.\\
We acknowledge the PCI BEV fellowship program from MCTI and CBPF.\\
This work is based on observations obtained with MegaPrime/MegaCam, a joint project of CFHT and CEA/DAPNIA, at the Canada--France--Hawaii Telescope (CFHT), which is operated by the National Research Council (NRC) of Canada, the Institut National des Sciences de l'Univers of the Centre National de la Recherche Scientifique (CNRS) of France, and the University of Hawaii. The Brazilian partnership on CFHT is managed by the Laborat\'orio Nacional de Astrof\'isica (LNA). We thank the support of the Laborat\'orio Interinstitucional de e-Astronomia (LIneA). We thank the CFHTLenS team for their pipeline development and verification upon which much of the CS82 survey pipeline was built.\\
Funding for SDSS-III has been provided by the Alfred P. Sloan Foundation, the Participating Institutions, the National Science Foundation, and the U.S. Department of Energy Office of Science. The SDSS-III web site is http://www.sdss3.org/.\\
SDSS-III is managed by the Astrophysical Research Consortium for the Participating Institutions of the SDSS-III Collaboration including the University of Arizona, the Brazilian Participation Group, Brookhaven National Laboratory, Carnegie Mellon University, University of Florida, the French Participation Group, the German Participation Group, Harvard University, the Instituto de Astrofisica de Canarias, the Michigan State/Notre Dame/JINA Participation Group, Johns Hopkins University, Lawrence Berkeley National Laboratory, Max Planck Institute for Astrophysics, Max Planck Institute for Extraterrestrial Physics, New Mexico State University, New York University, Ohio State University, Pennsylvania State University, University of Portsmouth, Princeton University, the Spanish Participation Group, University of Tokyo, University of Utah, Vanderbilt University, University of Virginia, University of Washington, and Yale University.\\
We made an extensively use of the following python libraries:  http://www.numpy.org/, http://www.scipy.org/, http://roban.github.com/CosmoloPy/ and http://www.matplotlib.org/.



\bibliographystyle{mnras}
\bibliography{references} 


\bsp	
\label{lastpage}
\end{document}